\documentclass[10pt,a4paper,aps,prb,reprint,nobibnotes,floatfix,superscriptaddress,showpacs]{revtex4-2}

\usepackage[utf8]{inputenc}
\usepackage[english]{babel}

\usepackage{amsfonts}
\usepackage{amsmath}
\usepackage{amssymb}
\usepackage[dvipsnames,svgnames,table]{xcolor}
\usepackage{graphicx}
\usepackage{fancyhdr}
\usepackage{float}
\usepackage{braket}
\usepackage{float}
\usepackage[caption=false]{subfig}
\usepackage{tikz}  % drawings
\usepackage{chngcntr}
\usepackage{ulem}

\usepackage{mathrsfs}

\usepackage[dvipsnames]{xcolor}

\usepackage{hyperref}
\hypersetup{
    pdfstartview={FitH},% fits the width of the page to the window
    colorlinks=true,    % false: boxed links; true: colored links
    linkcolor=NavyBlue, % color of internal links
    citecolor=Maroon,   % color of links to bibliography
    filecolor=NavyBlue, % color of file links
    urlcolor=NavyBlue   % color of external links
}

\makeatletter
\def\footnoterule{\kern -10pt
    \hrule \@width 100pt \kern 10pt} % the \hrule is .4pt high
\makeatother

\usepackage{tipa}

\renewcommand{\vec}[1]{\boldsymbol{#1}}
\newcommand{\dd}{\mathrm{d}}

\begin{document}

\title{Spin-Transfer Torque on Curved Surfaces: A Generalized Thiele Formalism}  

%%%%%%%%%%%%%%%%%%%%%%%%%%%%%%%%%%%%%%%%%%%%%%%%%%%%%%%%%%%%%%%%%%%%%
%%\listfiles
\def \udp {Universidad Diego Portales, Cedenna, Ej\'ercito 441, Santiago, Chile}
\def \fcfm {Departamento de F\'isica, FCFM, Universidad de Chile, Santiago, 8370448, Chile}

\def \brzil {Departamento de F\'isica, Universidade Federal de Vi\c cosa, Av. PH Rolfs s/n, 36570-900, Viçosa, Brazil}

\def \KravLeib {Leibniz-Institut f\"ur Festk\"orper- und Werkstoffforschung, Helmholtzstraße 20, D-01069 Dresden, Germany}

\def \KravBogo {Bogolyubov Institute for Theoretical Physics of the National Academy of Sciences of Ukraine, 03143 Kyiv, Ukraine}

\author{J. I. Costilla}
\affiliation{\brzil}
%\email{jose.pinedo@ufv.br}

\author{M. Castro}
\affiliation{\fcfm}
%\email{mario.castrob@usach.cl}

\author{K. V. Yershov}
\affiliation{\KravLeib}
\affiliation{\KravBogo}

\author{D. Altbir}
\affiliation{\udp}
%\email{dora.altbirb@udp.cl}

\author{V. L. Carvalho-Santos}
\affiliation{\brzil}
\email{vagson.santos@ufv.br}

\author{V. P. Kravchuk}
\affiliation{\KravLeib}
\affiliation{\KravBogo}

\begin{abstract}
Curvature is a highly relevant parameter when considering nanostructures, favoring the stability and affecting the dynamics of magnetic textures. In this work, we address the spin-transfer torque phenomenon by deriving an expanded Thiele equation with the Zhang-Li term for curved surfaces. Our results show a coupling between current and curvature, which is perceived as a gyrovector and an additional dissipative tensor associated with this coupling. Using this model, we determine the dynamics of a skyrmion in a nanotube with Gaussian and variable mean curvature. %Due to a curvature-induced force, the skyrmion propagates in the Walker regime. 
The new terms included in the Thiele equation are responsible for an additional Hall effect in the skyrmion dynamics and for the generalization of the Walker limit condition.
\end{abstract}

\maketitle
%%%%%%%%%%%%%%%%%%%%%%%%%%%%%%%%%%%%%%%%%%%%%%%%%%%%%%%%%%%%%%%%%%%%%

%%%%%%%%%%%%%%%%%%%%%%%%%%%%%%%%%%%%%%%%%%%%%%%%%%%%%%%%%%%%%%%%%%%%%

\textit{Introduction---} Interplay between the concepts of geometry and topology provides an important framework for understanding the properties of condensed matter systems. Geometry governs local physical properties and can induce effective interactions that do not appear in flat systems. Topology classifies the global structure of an order parameter in the so-called ordered media \cite{Mermin} and confers robustness through quantized invariants \cite{Chern-Simons}. The roles of geometry and topology are particularly important in magnetic systems. At small sizes, the shape of the magnetic structure is a key point in determining its properties, giving rise to the field of curvilinear micromagnetism \cite{Sheka2022,Sheka-APL,Makarov-Adv,Streubel,BookCurve}. This discipline explores the influence of geometry on the magnetization properties of magnetic bodies embedded in three-dimensional curved architectures. In these systems, the curvature of the underlying manifold acts as a tuning parameter, directly governing the topology of the magnetization field \cite{Example-1,Pacheco,Corona-Hopf,Castillo-Hopf, Kravchuk-PRL}. This geometry-topology interplay enables the stabilization of non-trivial magnetic textures and triggers unconventional static and dynamic responses in curved systems that are absent in their planar counterparts \cite{Gaididei-PRL,Yershov-Scipost,Kravchuk-PRB-Sk}.

The effects of geometry on ferromagnetic systems are directly evidenced by the influence of curvature on the behavior of magnetization collective modes. For example, the winding number of magnetic vortices is directly determined by the Gaussian curvature of a thin magnetic shell \cite{Example-2,Example-3,Beatriz-PRB,Elias2019}. Curvature also provides a mechanism for stabilizing skyrmions \cite{Kravchuk-PRB-Sk,Kravchuk-PRL,Tejo-ACS,Korniienko2020,Albertino}, while simultaneously affecting their intrinsic properties, such as size and shape \cite{Carvalho-JAP,Bichs-LTP,Yershov-SkDrift}. The influence of curvature extends to the dynamic properties of the magnetization by modifying the gyrovector field and the dissipative dyadic tensor  \cite{Korniienko2020}, which govern the motion of the soliton \cite{Thiele-PRL}. These curvature-induced changes give rise to different transport phenomena, including the emergence of pinning potentials \cite{Korniienko2020,CarvalhoSantos2021} and a transverse drift in the skyrmion trajectory \cite{Yershov-SkDrift,Coura-JMMM}.

Geometry effects extend to the current-driven dynamics of magnetic textures. In quasi-one-dimensional systems, for example, torsion and curvature modify the fundamental parameters of the spin transfer torque, effectively shifting the non-adiabatic coefficient \cite{Yershov2018,Bittencourt-NanoFut}. In systems with curvature gradients, the current-driven motion of a transverse domain wall breaks chiral symmetry, where the current threshold required for the domain wall to cross the bent region depends on its propagation direction \cite{Bittencourt-Nanoscale}. Although the influence of curvature on current-driven magnetic texture transport in one-dimensional structures has been addressed, its analysis in 2D magnetic systems has become a new area of research. To contribute to these challenges, in this letter, by deriving a generalized Thiele equation \cite{Thiaville05} for curved systems, we obtain a curvature-induced renormalization of both the gyrovector and the dissipative dyadic in the current-driven term. Using a geodesic approach, we apply this model to a skyrmion propagating along a bent nanotube. The curvature-induced changes in the spin-transfer torque generate transverse-to-current displacements of the skyrmion, culminating in a curvature-driven Magnus effect, even under equal Gilbert damping and non-adiabatic parameters, a regime that ensures linear trajectories in planar or cylindrical geometries. The analysis is extended to map the skyrmion dynamics on a nanotube, showing the emergence of a curvature-induced Walker-like regime for the skyrmion motion. 

\textit{Generalized Thiele formalism---} In the presence of spin-transfer torques (STT), the magnetization dynamics of a ferromagnet is governed by the Landau-Lifshitz-Gilbert equation  with additional Zhang-Li torques\cite{Zhang04}

\begin{equation}\label{eq:LL-ZhLi-main}
		\dot{\vec{m}}=\frac{\gamma}{M_s}\left[\vec{m}\times\frac{\delta \mathcal{H}}{\delta\vec{m}}\right]+\alpha\left[\vec{m}\times\dot{\vec{m}}\right]+\boldsymbol{\Lambda}_{\vec{u}},
\end{equation}

\noindent where a dot indicates the time derivative and $\boldsymbol{\Lambda_u}=\vec{m}\times\left[\vec{m}\times(\vec{u}\cdot\vec{\nabla})\vec{m}\right]+\beta \left[\vec{m}\times(\vec{u}\cdot\vec{\nabla})\vec{m}\right]$ is the Zhang-Li term due to the spin-transfer torque, where the velocity parameter $\vec{u}=-\vec{j}P\mu_\textsc{b}/(|e|M_s)$ is close to the drift velocity of the conductance electrons, $\beta$ is the non-adiabaticity constant,  $\vec{j}$ is the density of the current with the rate of polarization $P$, $\mu_\textsc{b}>0$ is the Bohr magneton, and $e$ is the electron charge.  Here, $\vec{m}=\vec{M}/M_s$ is the unit magnetization vector with $M_s$ being the saturation magnetization, $\gamma>0$ is the electron gyromagnetic ratio, and $\mathcal{H}=\int(\mathscr{E}_\text{ex} + \mathscr{E}_{\textsc{DM}} + \mathscr{E}_\text{an})\dd\vec{r}$ is the total magnetic Hamiltonian of the system. In addition, $\mathscr{E}_\text{ex}$, $\mathscr{E}_{\textsc{DM}}$, and $\mathscr{E}_\text{an}$ represent the energy densities for the exchange and Dzyaloshinskii-Moriya (DM) interactions, and the easy-normal anisotropy, respectively.  The parameter $\alpha$ is the Gilbert damping constant.  It is worth noticing that the form of Eq.~\eqref{eq:LL-ZhLi-main} is independent of the chosen reference frame.

\begin{figure}
\includegraphics[width=0.9\linewidth]{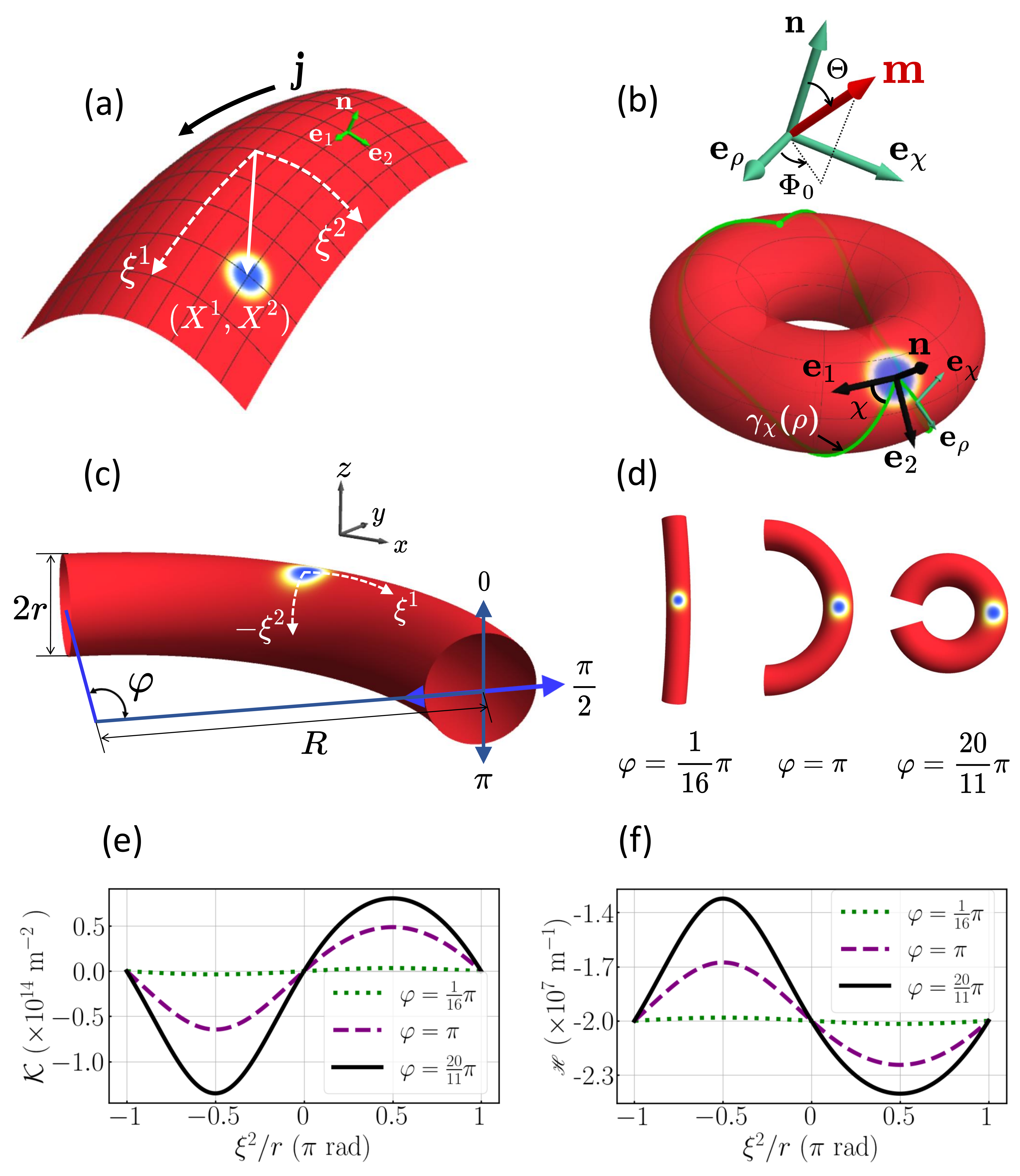}
\caption{Geometrical framework of the curved magnetic system. (a) Curvilinear coordinate system $(\xi^{1}, \xi^2)$  defining the manifold, with an inset showing the local magnetization parametrization via spherical angles. Coordinates $X^1$ and $X^2$ represent the position of the center of the skyrmion nucleated on the surface. (b) Diagram of a planar skyrmion profile mapped onto the curved surface. (c) Specific geometry of a bent nanotube. (d) Illustration of several nanotube opening angles and their resulting shape. (e,f) Spatial distribution of the Gaussian $\mathscr{K}$ and mean $\mathscr{H}$ curvatures for distinct opening angles with $r = 50 \ \text{nm}$, and $L=1130$ nm.}
\label{fig1}
\end{figure}

We now consider a very thin magnetic shell parametrized as $\boldsymbol{r}(\xi^1,\xi^2,\zeta)=\boldsymbol{\sigma}(\xi^1,\xi^2)+\zeta\,\boldsymbol{n}$, where $\vec{\sigma}(\xi^1,\xi^2)$ is the central surface of the shell with $\xi^1$ and $\xi^2$ being the curvilinear coordinates. Here, $\zeta \in [-\mathrm{h}/2, \mathrm{h}/2]$ is the coordinate through the thickness of the shell. We assume here that the surface thickness $\mathrm{h}$ is much smaller than the smallest curvature radius. The surface geometry can be characterized by the tangent basis vectors $\boldsymbol{g}_\mu=\partial \boldsymbol{\sigma} / \partial \xi^\mu$, and the associated metric tensor $g_{\mu\nu}=\boldsymbol{g}_\mu\cdot\boldsymbol{g}_\nu$, which describes the intrinsic geometry. The unit normal vector $\boldsymbol{n}\equiv\boldsymbol{e}_1\times\boldsymbol{e}_2$ defines the extrinsic orientation, as depicted in Fig.~\ref{fig1}(a). Here, $\boldsymbol{e}_\mu=\boldsymbol{g}_\mu/|\boldsymbol{g}_\mu|$ are the unitary vectors. In what follows, we assume that the basis $\{\vec{e}_1,\vec{e}_2\}$ is orthogonal and the metric tensor $g_{\mu\nu}$, accordingly, is diagonal.  

To take advantage of the convenience of representing the magnetization in a local reference frame, the unit magnetization vector can be parameterized as
$\boldsymbol{m}(\xi^1,\xi^2)=m^\alpha(\xi^1,\xi^2)\boldsymbol{e}_\alpha+m_n(\xi^1,\xi^2)\boldsymbol{n}$, where the Einstein sum convention (dumb repeated index represents sums) has been adopted. Here, we assume that the magnetization is uniform along the thickness, {\it i.e.}, it does not depend on $\zeta$. The STT $\vec{\Lambda}_{\vec{u}}$ are determined by the directional derivative $(\vec{u}\cdot\vec{\nabla})\vec{m}$. Assuming the tangential current $\vec{u}=u^\mu\vec{e}_\mu$, in the described curvilinear reference frame, we obtain
$(\vec{u}\cdot\vec{\nabla})\vec{m}=u^\mu[\vec{e}_\nu(\nabla_\mu {m}^\nu+\Omega_\mu\epsilon_{\nu\lambda}{m}^\lambda-h_{\mu\nu}{m}_n)+\vec{n}(\nabla_\mu {m}_n+h_{\mu\nu}{m}^\nu)]+\mathcal{O}(\zeta)$. Here, $\nabla_\mu=g_{\mu\mu}^{-1/2}\partial_\mu$ is the gradient component, $\Omega_\mu=\vec{e}_1\cdot\nabla_\mu\vec{e}_2$ is the spin connection, and $h_{\mu\nu}=\vec{n}\cdot\nabla_\mu\vec{e}_\nu$ is the shape operator also known as Weingarten map. The terms proportional to the shape operator elements generate additional driving forces emerging from the interplay between spin-transfer torques (STT) and curvature. Within the same framework, we highlight that one obtains a curvature-induced DMI and an effective anisotropy,
which are linear and quadratic in $h_{\mu\nu}$, respectively~\cite{Gaididei-PRL,Sheka15,Sheka22b}. The eigenvalues of $h_{\mu\nu}$ are the principal curvatures $\kappa_1$ and $\kappa_2$ that determine the mean $\mathscr{H}=\kappa_1+\kappa_2$ and the Gaussian $\mathscr{K}=\kappa_1\kappa_2$ curvatures. 

Let us now assume that in the curvilinear reference frame $(\xi^1,\xi^2)$, the dynamics of the magnetic texture can be presented in the form of a traveling-wave Ansatz $\vec{m}=\mathrm{m}^\mu(\xi^1-X^1,\xi^2-X^2)\vec{e}_\mu+\mathrm{m}_n(\xi^1-X^1,\xi^2-X^2)\vec{n}$, where $\mathrm{m}^\mu(\xi^1,\xi^2)$ and $\mathrm{m}_n(\xi^1,\xi^2)$ are known functions (e.g., the Ansatz for the skyrmion profile) and the time dependence is included in the collective coordinates $X^\mu=X^\mu(t)$, which determine the position of the magnetic pattern. Using this formalism, in the leading order in thickness $\mathrm{h}$ and in the linear order in principal curvatures, the extended Thiele equation including the Zhang-Li torque for the collective coordinates in a curved background is derived from Eq. \eqref{eq:LL-ZhLi-main}, and is given by 

\begin{equation}\label{eq:Thiele-Main}
G_{\mu\nu}(V^\nu-u^\nu)=-F_\mu+D_{\mu\nu}(\alpha V^\nu-\beta u^\nu)+(G^u_{\mu\nu}-\beta D^u_{\mu\nu})u^\nu,
\end{equation}

\noindent whose derivation is presented in the Supplemental Information (SI) \cite{SupMat}. Here, $V^\mu=\sqrt{g_{\nu\nu}(X^1,X^2)}\dot{X}^\mu$ are components of the skyrmion velocity, and $F_\mu=-[g_{\mu\mu}(X^1,X^2)]^{-\frac12}\partial\mathcal{H}/\partial X^\mu$ is the curvature-induced force (CIF) acting on the skyrmion. The gyrotensor $G_{ab}$ and the damping tensor $D_{\mu\nu}$ were previously obtained for curved films~\cite{Korniienko2020} hosting a skyrmion, showing that, at the limit where the curvature radii are much larger than the skyrmion size, $G_{\mu\nu}$ and $D_{ab}$ tend towards the values of a flat film~\cite{Korniienko2020}. The central finding of this work is the derivation of a generalized framework from which emerge two curvature-mediated current-driven forces: a current-curvature-induced gyrotensor (CCG, $G^u_{\mu\nu}$) and a dissipative dyadic (CCD, $D^u_{\mu\nu}$). Their tensorial structure is directly governed by the background curvature, revealing a fundamental, hitherto unexplored coupling between spin currents, magnetization dynamics, and curvature. Both the CCG and CCD vanish in the case of a flat surface, so that Equation~\eqref{eq:Thiele-Main} naturally reduces to the generalized Thiele equation for a flat stripe~\cite{Thiaville05}.

To compute the coefficients in the generalized Thiele equation \eqref{eq:Thiele-Main}, we write the skyrmion Ansatz in geodesic polar coordinates (GPC) (see SI \cite{SupMat} for details). For this purpose, from the center of the skyrmion $(X^1,X^2)$, we launch geodesics $\vec{\gamma}_\chi(\rho)$ at different angles $\chi$ relative to the principal direction $\vec{e}_1$, where $\rho$ is the geodesic arc-length starting from the center of the skyrmion and $\vec{e}_\rho=\partial_\rho\vec{\gamma}_\chi$ is the unit vector tangential to $\vec{\gamma}_\chi$. Using the orthonormal coordinate-dependent basis $\{\vec{e}_\rho,\vec{e}_\chi,\vec{n}\}$, with $\vec{e}_\chi=\vec{n}\times\vec{e}_\rho$, one can formulate the skyrmion Ansatz $\mathbf{m}=\sin\Theta(\rho)[\vec{e}_\rho\cos\Phi_0+\vec{e}_\chi\sin\Phi_0]+\cos\Theta(\rho)\vec{n}$. Here, $\Theta(\rho)$ is the skyrmion profile in a planar film where the constant $\Phi_0$ determines the skyrmion type. That is,  $\Phi_0=0,\pi$ and $\Phi_0=\pm\pi/2$ correspond to a N{\'e}el and a Bloch skyrmion, respectively. The Ansatz used here assumes that $\vec{\mathrm{m}}=\pm\vec{n}$ at a large distance from the center of the skyrmion. The latter means that we neglect the curvature-induced deviation of the ground state from the normal direction, which is generally~\footnote{This deviation appears not for all surfaces.} of the order $\mathcal{O}(\ell^2\kappa^2_\alpha,\ell^2\kappa_1\kappa_2)$~\cite{Gaididei-PRL}. Here $\ell=\sqrt{A_{ex}/K}$ is the exchange length with $A_{ex}$ and $K$ being the exchange stiffness and the coefficient of the easy-normal anisotropy, respectively.

Using the GPC-Ansatz, in the linear order of the principal curvatures, we obtain the gyro- and damping-tensors as in a planar film, namely $G_{\mu\nu}=\epsilon_{\mu\nu}G$, with $G=4\pi N_{\text{top}}M_s\mathrm{h}/\gamma$, and $D_{\mu\nu}=\delta_{\mu\nu}D$, with $D=4\pi\mathcal{C}_0M_s\mathrm{h}/\gamma$. Here, $N_{\mathrm{top}}=\frac12\left[\cos\Theta(0)-\cos\Theta(\infty)\right]=\pm1$ is the skyrmion topological charge, and the dimensionless constant $\mathcal{C}_0=\frac{1}{4}\int_0^\infty[\Theta'^2+\rho^{-2}\sin^2\Theta]\rho\dd\rho$ is determined by the skyrmion profile. The CCD is given by
\begin{equation}
D_{\mu\nu}^u=-4\pi\frac{\mathrm{h}M_s}{\gamma}\mathcal{C}_1 \hat{R}_{\mu\lambda}h_{\lambda\nu}
\end{equation}
where $\hat{R}_{\mu\nu}=\delta_{\mu\nu}\cos\Phi_0+\epsilon_{\mu\nu}\sin\Phi_0$ is the rotation matrix, $h_{\mu\nu}=h_{\mu\nu}(X^1,X^2)$ are elements of the shape operator in the skyrmion center, and $\mathcal{C}_1=\frac{1}{4}\int_0^\infty[\Theta'+\rho^{-1}\sin\Theta\cos\Theta]\rho\dd\rho$. 
Within the framework of the approximations made, we obtain $G^u_{\mu\nu}=0$. However, taking into account the terms of higher order in the principal curvatures and the curvature-induced deformation of the skyrmion profile can lead to non-zero elements of the tensor $G^u_{\mu\nu}$.\footnote{Note that in this case, Thiele equation \eqref{eq:Thiele-Main} can have a more general form, as presented in Eq.(S.5) of SI \cite{SupMat}.} For example, considering a cylindrical surface of general form with $[h_{\mu\nu}]=\text{diag}(\kappa_1(\xi^1),0)$ and $\Gamma_{\mu\nu}^\lambda=0$, we take into account the skyrmion deformation as described in Ref.~\onlinecite {Yershov-SkDrift} and find that the leading order correction to $G^u_{\mu\nu}$ is proportional to $\ell^3\kappa''_1(X^1)$.\footnote{We obtain $G^u_{11}=-\pi\sin\Phi_0\kappa''_1(X^1)\mathcal{C}_3\mathrm{h}M_s/\gamma$, where $\mathcal{C}_3=\int_0^\infty\dd\rho\rho^2\sin\Theta(\rho)$, and the other components of the tensor $G^u_{\mu\nu}$ are of the higher order in curvature.}

We now compute the curvature-induced correction of the skyrmion energy for the case of interfacial DMI: $    \Delta E_{\text{N{\'e}el}}=-8\pi \mathrm{h}(\cos\Phi_0 A_{ex}\mathcal{C}_1-\mathcal{D}_{\textsc{dm}}\mathcal{C}_2) \mathscr{H}(X^1,X^2)+\mathcal{O}(\kappa^2_1,\kappa_2^2,\kappa_1\kappa_2)$ (See details in SI), where, $\mathcal{D}_{\textsc{dm}}$ is the DMI constant,  and $\mathcal{C}_2=\frac14\int_0^\infty\sin^2\Theta\rho\dd\rho$.  Thus, we reproduce the results previously obtained~\cite{Korniienko2020,Yershov-SkDrift} for a coarser Ansatz of a skyrmion on a curved surface.  For an N{\'e}el skyrmion, the flip of the sign of $\mathcal{D}_{\textsc{dm}}$ flips the sign of $\cos\Phi_0$, and therefore flips the sign of $\Delta E_{\text{N{\'e}el}}$. At the same time, flipping the ground state direction, which is assumed parallel or antiparallel to $\vec{n}$, does not change $\Delta E_{\text{N{\'e}el}}$, because both $\cos\Phi_0$ and $\mathcal{C}_1$ flip signs in this case.

For a Bloch skyrmion, the curvature-dependent energy correction $\Delta E_{\text{Bloch}}=\mathcal{O}(\kappa^2_1,\kappa_2^2,\kappa_1\kappa_2)$ is of the second order in the principal curvatures. Note that the theory developed here takes into account only corrections linear in the curvature; therefore, generally it can not be applied to Bloch skyrmions.

\textit{Toroidal surface as a case study---}To illustrate new effects stemming from the current-curvature-induced terms in Eq.~\eqref{eq:Thiele-Main}, we consider the geometry of a bent tube, which is modeled as a section of a torus with length $L$, an opening angle $\varphi$, a toroidal radius $R=L/\varphi$, and a tube radius $r$. The tube can be parametrized as $\boldsymbol{\sigma}(\xi^1,\xi^2) = [R + r\sin(\xi^2/r)]\sin(\xi^1/R)\mathbf{\hat{x}} + [R + r\sin(\xi^2/r)]\cos(\xi^1/R)\mathbf{\hat{y}} + r\cos(\xi^2/r)\mathbf{\hat{z}}$. Here, $\xi^1\in[0,2\pi R]$ and $\xi^2\in[-\pi r,\pi r]$ are the curvilinear coordinates associated with the toroidal and poloidal directions, respectively, as shown in Figs. \ref{fig1}(c) and (d). The chosen parameterization results in the diagonal shape operator $[h_{\mu\nu}]=\text{diag}(\kappa_1,\kappa_2)$ with $\kappa_1=-R^{-1}\sin\vartheta/(1+\varrho\sin\vartheta)$ and $\kappa_2=-r^{-1}$ being the principal curvatures. Here $\vartheta=\xi^2/r$ is the poloidal angle and $\varrho=r/R$ defines the aspect ratio between the tube and toroidal radii.

\begin{figure}
\includegraphics[width=1.0\linewidth]{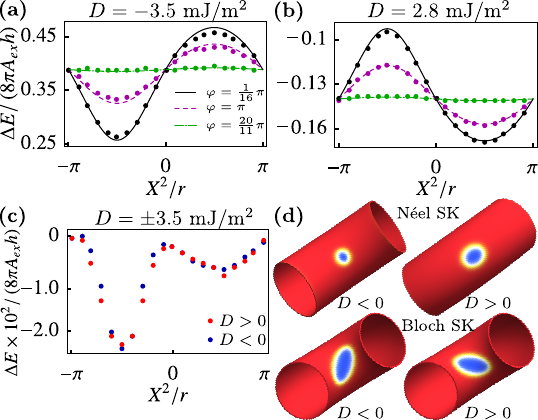}
\caption{Skyrmion energy as a function of its position. (a) and (b) show the energy of the N{\'e}el skyrmion for different opening angles and $L=1130$ nm, (c) shows the energy of the Bloch skyrmion for a fixed opening angle $\varphi = 20\pi/11$ and $L=565$ nm. (d) Shows the skyrmion profile reconstructed from micromagnetic simulations\cite{Note3} in its equilibrium position: top and bottom rows correspond to the N{\'e}el and Bloch skyrmions, respectively, for negative and positive DMI values. Symbols in (a)-(c) correspond to the data obtained by means of numerical simulations\cite{Note3}, and lines in (a) and (b) correspond to the analytical predictions.} 
\label{fig2}
\end{figure}

In this work, we focus on the bending-induced effects in the current-driven skyrmion dynamics, absent in the straight cylinder~\cite{Wang-PRB-2023,Wang2019}. We begin by analyzing the curvature-induced energy potential. For the chosen geometry, the curvatures depend solely on the poloidal coordinate $\xi^2$, see Fig.~\ref{fig1}(e,f). As a result, the curvature-induced potential depends only on the skyrmion coordinate $X^2$. In Fig.~\ref{fig2}(a,b), we compare our analytical prediction $\Delta E_{\text{N{\'e}el}}$ for the Néel skyrmion to the values extracted from micromagnetic simulations. Our results show a good agreement between the numerical data and our estimations for $\Delta E_{\text{N{\'e}el}}\propto\mathscr{H}$ obtained in the linear order in $\kappa_\alpha$. Thus, the minimum energy is achieved when the N{\'e}el skyrmion is on the outer ($X^2=\frac{\pi}{2}r$) or inner ($X^2=-\frac{\pi}{2}r$) part of the torus surface for $\mathcal{D}_{\textsc{dm}}>0$ or $\mathcal{D}_{\textsc{dm}}<0$, respectively.

Since at this stage our theory does not allow to take into account second-order corrections in $\kappa_\mu$, for Bloch skyrmions we present only numerical data extracted from the simulations~\footnote{All simulations have the following common parameters: $r=50$~nm, $h=2$ nm, $A_{ex}=16\times 10^{-12}$ J/m, $K=5.6\times 10^5$ J/m$^3$, $M_s = 1.1\times10^6$ A/m.}, see Fig.~\ref{fig2}(c). In contrast to the N{\'e}el skyrmion, we observe that the energy landscape for the Bloch skyrmion is not affected by the sign of $\mathcal{D}_{\textsc{dm}}$. A Bloch skyrmion has two equilibrium positions: $X^2=\pm\frac{\pi}{2}r$. However, the global minimum corresponds to the inner position $X^2=-\frac{\pi}{2}r$,  where the Gaussian curvature reaches its extreme negative value, as shown in Fig.~\ref{fig1}(e). This later is intrinsically related to the winding-number selection of solitonic magnetic textures in curved systems \cite{Vitelli,Example-2,Example-3,Beatriz-PRB} and to curvature-induced magnetochirality \cite{Yershov-2015,Volkov-PRL-2019}. Note that $N_{\text{top}}=-1$ for the considered skyrmions. Fig.~\ref{fig2}(d) shows the images of  N{\'e}el and Bloch skyrmions relaxed in their equilibrium positions. For the Bloch skyrmion, the amplitude of the energy landscape is two orders of magnitude lower than that of the N{\'e}el skyrmion. To extract the curvature-induced energy of the Bloch skyrmion from the simulations, we significantly increase the curvature. The latter results in the noticeable elliptical deformation of the skyrmion shape~\cite{Bichs-LTP,Yershov-SkDrift}, as illustrated in Fig.~\ref{fig2}(d). Although the rigid skyrmion Ansatz remains valid for Bloch skyrmions under certain constraints, in the following, we restrict our analytical model exclusively to Néel skyrmions.

For the torus, the curvature-induced force $\vec{F}=F_2\vec{e}_2$ has only one component $F_2=8\pi\mathrm{h}\Xi r^{-1}\partial_\vartheta\kappa_1$, where $\Xi=A_{ex}\mathcal{C}_1-\mathcal{D}_{\textsc{dm}}\mathcal{C}_2$. Here and in what follows, we address the case $\mathcal{D}_{\textsc{dm}}>0$ and $\cos\Phi_0=1$. Note that constants $\Xi$ and $\mathcal{C}_n$ are uniquely determined by the dimensionless DMI constant $d=\mathcal{D}_{\textsc{dm}}/\sqrt{A_{ex}K}$ or by skyrmion radius (see Fig.~S3 and Ref. [\onlinecite{Yershov-SkDrift}]). We will apply the current in the toroidal direction $\vec{u}=u^1\vec{e}_1$, which is technically accessible and convenient for the analysis. In this case, the Thiele Eqs.~\eqref{eq:Thiele-Main} can have a steady-state solution $V^1=u^1+F_2(\vartheta_0)/G$, $V^2=0$, where the poloidal angle of the skyrmion position $\vartheta_0=X^2_0/r$ is determined by the equation
\begin{equation}\label{eq:Xc}
    \alpha F_2(\vartheta_0)=Gu^1\left[\beta\frac{D^u_{11}(\vartheta_0)}{D}+\beta-\alpha\right].
\end{equation}
For the simplest hypothetical case $u^1=0$ and $\alpha=0$, Eq.~\eqref{eq:Xc} is turned into an identity and the skyrmion velocity $V^1=F_2(\vartheta_0)/G$ is determined solely by the initial skyrmion coordinate $\vartheta_0$, which can be arbitrary. The frequency of the toroidal skyrmion motion is $\Omega=\Omega_0\cos\vartheta_0/(1+\varrho\sin\vartheta_0)^3$, where $\Omega_0=-\frac{\gamma\Xi}{N_{\text{top}}M_s}\frac{1}{2\pi R^2r}$. In the limit $\varrho\ll1$, the frequency value reaches its maximal values $\Omega_{\text{max}}\approx\Omega_0$ and $\Omega_{\text{max}}\approx-\Omega_0$ for $\vartheta_0\approx-3\varrho$ and $\vartheta_0\approx-\pi+3\varrho$, respectively. This corresponds to the almost extreme top or bottom skyrmion positions, with a tiny shift towards the inner part of the torus.

\begin{figure}
    \centering
    \includegraphics[width=1.0\linewidth]{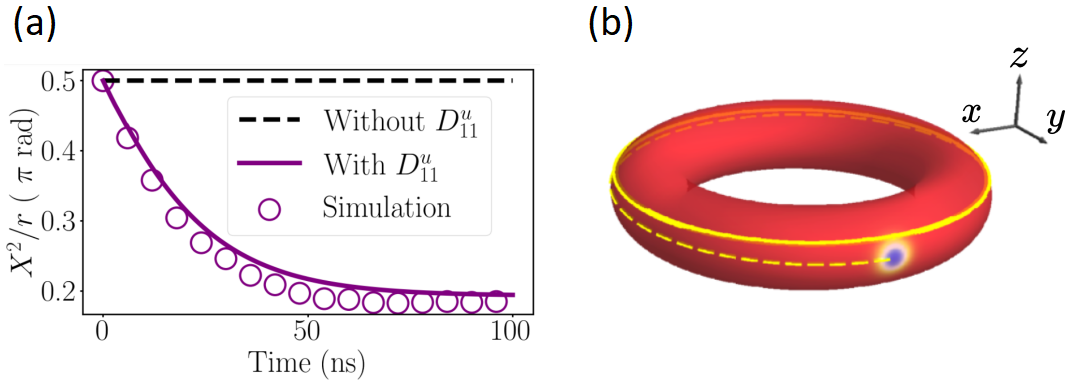}
    \caption{Skyrmion trajectory for $\alpha=\beta = 0.5$, $|J|=10\times10^{12}$ A/m $^2$, $\varphi=2\pi$, and $L=1130$~nm. (a) Skyrmion position as a function of time. Purple line and circle depict the trajectory predicted by the analytical model (line) and micromagnetic simulations\cite{Note3} (circles). The black-dashed line shows the skyrmion trajectory in the absence of the CCD. (b) depicts the skyrmion trajectory along the bent tube. The dashed line shows the trajectory during the first 50 ns.}
    \label{Fig4}
\end{figure}

For the particular case $\alpha=\beta$, Eq.~\eqref{eq:Xc} is reduced to $\cot\vartheta_0/(1+\varrho\sin\vartheta_0)=u^1/u_0$, which has a solution for any current,  $u^1$, value. Here $u_0=4\pi\Omega_0\mathcal{C}_0R^2/\mathcal{C}_1$. In the limit $\varrho\ll1$ (thin torus), the skyrmion velocity is
\begin{equation}
    V^1\approx u^1\left[1+\frac{\mathcal{C}_1}{2\mathcal{C}_0R}\frac{1}{\sqrt{1+(u^1/u_0)^2}}\right].
\end{equation}
In the limit of large currents $|u^1|\gg|u_0|$, one obtains $V^1\approx u^1+V_g$, where $V_g=2\pi R\Omega_0$ is the correction stemming from the geometrically induced potential. As the current increases, the skyrmion reaches the top ($\vartheta_0\to0$) or bottom ($\vartheta_0\to\pm\pi$) trajectory. It depends on the signs of $u^1$, $u_0$, as well as on the initial skyrmion position. The described skyrmion dynamics is essentially determined by the new term $D^u_{11}$; its crucial role is illustrated in Fig.~\ref{Fig4}.
In a flat system, the condition $\alpha=\beta$ suppresses the transverse motion from the standard skyrmion Hall effect in a flat system  \cite{Ohki_2025}. %However, in our case, CCD leads to a transversal velocity component, as shown in the dashed trajectory in Fig.~\ref{Fig4}(b).
However, in our case, there is a velocity component perpendicular to the current direction. The skyrmion's transverse velocity $\dot{X}^2$ vanishes only at a new steady position. Results for a nanotube with $r=50$ nm, $L=1130$ nm, and $\varphi=2\pi$ are shown in Fig. \ref{Fig4}. The skyrmion starts its motion at $X^2/r=0.5$ and propagates toward a new steady state, at $X^2/r\approx0.2$, corresponding to a distance of 15 nm from the initial position. The steady state is determined by the balance between the CIF and the effective force originated from the CCD term. During the transient propagation, the skyrmion exhibits a velocity transverse to the current direction, as evidenced by its poloidal displacement represented by the red line shown in Fig. \ref{Fig4}(a). The skyrmion position in time is sketched in the schematic representation given in Fig. \ref{Fig4}(b). Without current-curvature effects this initial transverse motion would not occur, as shown by the black-dashed line in Fig. \ref{Fig4}(a). To validate the analytical findings, we have performed micromagnetic simulations using the TetMag code \cite{TetMag} under identical initial conditions (details of simulations are given in SI). Our results, shown by purple circles, confirm the emergence of an extra Hall effect stemming from current-curvature coupling in curved geometries.

\begin{figure}
    \centering
    \includegraphics[width=\columnwidth]{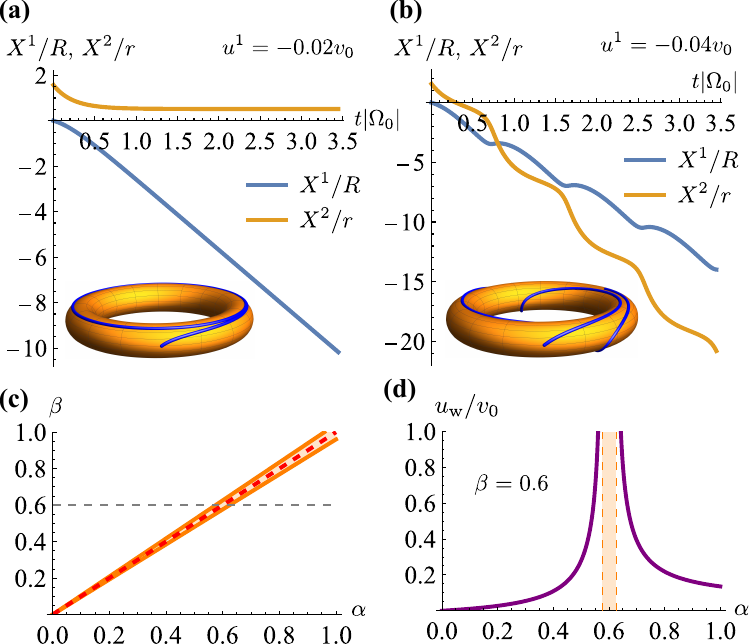}
    \caption{Comparison of the translational and oscillatory skyrmion motions is shown in panels (a) and (b), respectively. Here, we consider a torus with $R=10\ell$ and $r=2\ell$, and a skyrmion with $N_{\text{top}}=-1$ with radius $R_s=0.5\ell$. These parameters correspond to $V_g\approx-0.027v_0$ with $v_0=\gamma\sqrt{A_{ex}K}/M_s$, and $\Omega_0\approx-4.3\times10^{-4}\gamma K/M_s$. For $\alpha=0.1$ and $\beta=0.3$ we obtain $\Upsilon\approx0.062$ and estimate from \eqref{eq:uw} $u_{\text{w}}\approx0.0271v_0$. The exact value numerically obtained from \eqref{eq:Thiele-Main} is $u_{\text{w}}\approx0.0286v_0$. For the considered values of $R$, $r$, and $R_s$, the condition $\Upsilon>1$ is shown on panels (c) and (d) by the orange shadowing. The dependence $u_{\mathrm{w}}(\alpha)$ determined by Eq.~\eqref{eq:uw} is shown on panel (d) for fixed $\beta=0.6$.}
    \label{fig:trajes}
\end{figure}

For the general case $\alpha\ne\beta$, Eq.~\eqref{eq:Xc} can only be resolved for certain parameter values, meaning that a translational skyrmion motion is not always possible. Instead, a skyrmion can experience an oscillatory dynamics, during which it performs periodic rotations in the poloidal direction.  The condition for the transition from one type of dynamics to another is more complex than the conventional Walker breakdown~\cite{Thiaville05,Yershov-2016} induced by the spin-transfer torque. However, it can be easily analyzed in the limit $\varrho\ll1$. In this limit, the translational motion occurs for any value of the applied current, if  $\Upsilon=\frac{\beta}{|\alpha-\beta|}\frac{|\mathcal{C}_1|}{\mathcal{C}_0R}>1$. For $\Upsilon<1$, the translational motion occurs only if the current does not exceed some critical value (Walker limit): $|u^1|<u_{\text{w}}$, where
\begin{equation}\label{eq:uw}
    u_\text{w}=2|V_g|\frac{\alpha}{|\alpha-\beta|}\frac{1}{\sqrt{1-\Upsilon^2}}.
\end{equation}
The translational and the oscillatory motions are compared in Fig.~\ref{fig:trajes}(a,b).
For a straight cylinder ($R\to\infty$), one has $V_g=0$,  $\Upsilon=0$ and consequently $u_{\text{w}}=0$. In this case, the translational motion occurs only under the condition $\alpha=\beta$. This is consistent with previous results~\cite{Wang-PRB-2023,Wang2019}. A direct consequence of the new CCD term $D^u_{11}$ is the change in the condition of a guaranteed translational motion, namely $\Upsilon>1$, instead of the condition $\alpha=\beta$ relevant for the planar systems~\cite{Thiaville05}. Within the plane $(\alpha,\beta)$, the condition $\Upsilon>1$ determines a 2D area surrounding the line $\alpha=\beta$, as illustrated in Fig.~\ref{fig:trajes}(c). The critical velocity $u_{\mathrm{w}}$ becomes infinitely large at the boundary of the region determined by the condition $|\Upsilon|=1$, see Fig.~\ref{fig:trajes}(d). For the case of one-dimensional skyrmion motion along the toroidal coordinate, the action of the $D^u_{\mu\nu}$ is equivalent to a rescaling of the nonadiabatic parameter $\beta_{\text{eff}}=\beta(1+D^u_{11}/D)$. Note that $\beta_{\text{eff}}=\beta_{\text{eff}}(\vartheta_0)$ depends on the skyrmion position. The condition $\Upsilon>1$ means that there exists a trajectory $\vartheta=\vartheta_0$ such that the condition $\alpha=\beta_{\text{eff}}(\vartheta_0)$ holds. Note that the modification of the Walker limit \eqref{eq:uw} is valid for the toroidal geometry, and it can have a different form for the other geometries.

\textit{Conclusions.---} We generalized the Thiele equation for the case of the current-driven skyrmion dynamics along a thin curvilinear film. We demonstrated that curvature-induced modification of the spin-transfer torques results in two additional terms in the Thiele equation, namely $D_{\mu\nu}^u$ and $G^u_{\mu\nu}$ tensors.  Within the approximations used (linear order in the principal curvatures and neglecting curvature-induced skyrmion deformation), tensor $G_{\mu\nu}^u$ vanishes. However, its presence may be important for more accurate models of skyrmions or for the Thiele dynamics of other topological solitons. As an example, we considered the motion of a skyrmion along a toroidal surface and showed that the presence of the tensor $D_{\mu\nu}^u$ extends the condition of the Walker limit and leads to the additional Hall effect in the skyrmion dynamics. It is worth mentioning that we have also developed a new framework to describe the skyrmion on a curved surface, based on the geodesic polar coordinates. 

\textit{Acknowledgments.---} In Brazil, we thank Capes (Grant N. 001) and INCT/CNPq - Spintr{{\^o}nica e Nanoestruturas Magn\'eticas Avan\c{c}adas (INCT-SpinNanoMag), CNPq 406836/2022-1. In Chile we acknowledge CEDENNA under grant CIA250002 from ANID.   M. Castro acknowledges Proyecto ANID Fondecyt Postdoctorado 3240112. V.K. acknowledges financial support by the Deutsche Forschungsgemeinschaft (DFG, German Research Foundation) through the W{\"u}rzburg-Dresden Cluster of Excellence ctd.qmat – Complexity, Topology and Dynamics in Quantum Matter (EXC 2147, project-id 390858490). K.Y. acknowledges financial support from the Deutsche Forschungsgemeinschaft (DFG, German Research Foundation)
under grant No. YE 232/2-1.

\textit{Data Availability.---} The data are available from the authors upon reasonable request.

\bibliography{Bibliography}

\clearpage
\break
\onecolumngrid

\begin{center}
	\textbf{\large SUPPLEMENTARY MATERIALS}
\end{center}
%	\beginsupplement
\renewcommand{\thefigure}{S\arabic{figure}}
\renewcommand{\theequation}{S.\arabic{equation}}

    In these supplemental materials, we provide details of the derivation of the generalized Thiele equation for a thin curvilinear film with spin-transfer torques taken into account. The equation obtained is applied to describe the current-driven dynamics of a skyrmion on a toroidal surface.

    \section{Generalized Thiele equation} 

Let us start by considering a general model of collective variables $\vec{m}(t,\vec{r})=\mathbf{m}(\vec{r};X^1(t),X^2(t),\dots)$, in which the time dependence is included in parameters $X^i(t)$ of arbitrary meaning. Taking into account that $\dot{\mathbf{m}}=\frac{\partial\mathbf{m}}{\partial X^j}\dot{X}^j$, we apply the operation $\int\mathbf{m}\cdot[\frac{\partial\mathbf{m}}{\partial X^i}\times\dots]\dd\vec{r}$ to the Landau-Lifshits-Gilbert equation
\begin{equation}\label{eq:LL-ZhLi}
		\dot{\vec{m}}=\frac{\gamma}{M_s}\left[\vec{m}\times\frac{\delta \mathcal{H}}{\delta\vec{m}}\right]+\alpha\left[\vec{m}\times\dot{\vec{m}}\right]+\vec{m}\times\left[\vec{m}\times(\vec{u}\cdot\vec{\nabla})\vec{m}\right]+\beta \left[\vec{m}\times(\vec{u}\cdot\vec{\nabla})\vec{m}\right].
\end{equation}
When considering a thin film with uniform magnetization along the thickness (independent of $\zeta$), in the leading order in thickness $h$, we obtain the equation for the collective variables:
\begin{equation}\label{eq:Thiele-gen}
    \mathcal{G}_{ij}\dot{X}^j=\frac{\partial\mathcal{H}}{\partial X^i}+\alpha \mathcal{D}_{ij}\dot{X}^j+\tilde{\mathcal{J}}^{(1)}_i+\beta\tilde{\mathcal{J}}^{(2)}_i.
\end{equation}
Here,
\begin{equation}\label{eq:CV-coefs}
    \begin{split}
        &\mathcal{G}_{ij}=h\frac{M_s}{\gamma}\int\mathbf{m}\cdot\left[\frac{\partial\mathbf{m}}{\partial X^i}\times\frac{\partial\mathbf{m}}{\partial X^j}\right]\dd S,\qquad
        \mathcal{D}_{ij}=h\frac{M_s}{\gamma}\int\frac{\partial\mathbf{m}}{\partial X^i}\cdot\frac{\partial\mathbf{m}}{\partial X^j}\dd S,\\
        &\tilde{\mathcal{J}}^{(1)}_i=-h\frac{M_s}{\gamma}\int\mathbf{m}\cdot\left[\frac{\partial\mathbf{m}}{\partial X^i}\times\left(\vec{u}\cdot\vec{\nabla}\right)\mathbf{m}\right]\dd S,\qquad\tilde{\mathcal{J}}^{(2)}_i=h\frac{M_s}{\gamma}\int\frac{\partial\mathbf{m}}{\partial X^i}\cdot\left[\left(\vec{u}\cdot\vec{\nabla}\right)\mathbf{m}\right]\dd S \,
    \end{split}
\end{equation}
where $\dd S=\sqrt{|g|}\dd\xi^1\dd\xi^2$ is the element of the surface area and $g=\det(g_{\alpha\beta})$ is the determinant of the metric tensor. Let us now consider the particular case in which the collective variables are collective coordinates, {\it i.e.} $\vec{m}=\mathrm{m}^\alpha(\xi^1-X^2,\xi^2-X^2)\vec{e}_\alpha+\mathrm{m}_n(\xi^1-X^2,\xi^2-X^2)\vec{n}$, where $\{\vec{e}_1,\vec{e}_2,\vec{n}\}$ is an orthonormal basis such that $\vec{e}_\alpha=\vec{g}_\alpha/\sqrt{g_{\alpha\alpha}}$ and $\vec{n}=\vec{e}_1\times\vec{e}_2$. Here, $\vec{g}_\alpha=\partial_\alpha\vec
\sigma(\xi^1,\xi^2)$ is a covariant basis tangential to the surface $\vec{\sigma}=\vec{\sigma}(\xi^1,\xi^2)$. For this case $\frac{\partial\mathbf{m}}{\partial X^\alpha}=\vec{e}_{\beta}\frac{\partial\mathrm{m}^\beta}{\partial X^\alpha}+\vec{n}\frac{\partial m_n}{\partial X^\alpha}=-\vec{e}_\beta\partial_\alpha \mathrm{m}^\beta-\vec{n}\partial_\alpha\mathrm{m}_n$. Note that the basis vectors $\vec{e}_\alpha$ and $\vec{n}$ do not depend on the collective variables $X^\alpha$. A current flowing along the surface can be described as $\vec{u}=u^\alpha\vec{e}_\alpha$. Taking also into account that $(\vec{u}\cdot\vec{\nabla})\mathbf{m}=u^\alpha[\vec{e}_\beta(\nabla_\alpha \mathrm{m}^\beta+\Omega_\alpha\epsilon_{\beta\gamma}\mathrm{m}^\gamma-h_{\alpha\beta} \mathrm{m}_n)+\vec{n}(\nabla_\alpha \mathrm{m}_n+h_{\alpha\beta}\mathrm{m}^\beta)]$~\footnote{We used here the Gauss formula $\nabla_\alpha\vec{e}_\beta=h_{\alpha\beta}\vec{n}-\Omega_\alpha\epsilon_{\beta\gamma}\vec{e}_\gamma$ and the Weingarten formula $\nabla_\alpha\vec{n}=-h_{\alpha\beta}\vec{e}_\beta$.}, with $\nabla_\alpha=g_{\alpha\alpha}^{-1/2}\partial_\alpha$ being the gradient components, $\Omega_\alpha=\vec{e}_1\cdot\nabla_\alpha\vec{e}_2$ is the spin connection, and $h_{\alpha\beta}=\vec{n}\cdot\nabla_\alpha\vec{e}_\beta$ is the shape operator (a.k.a. Weingarten map), we can write \eqref{eq:CV-coefs} in the following explicit form
\begin{equation}\label{eq:coefs-Thiele}
    \begin{split}
        &\mathcal{G}_{\alpha\beta}=h\frac{M_s}{\gamma}\int\begin{vmatrix}
            \mathrm{m}^1 & \partial_\alpha\mathrm{m}^1 & \partial_\beta\mathrm{m}^1 \\
            \mathrm{m}^2 & \partial_\alpha\mathrm{m}^2 & \partial_\beta\mathrm{m}^2 \\
            \mathrm{m}_n & \partial_\alpha\mathrm{m}_n & \partial_\beta\mathrm{m}_n
        \end{vmatrix}\dd S,\qquad \mathcal{D}_{\alpha\beta}=h\frac{M_s}{\gamma}\int\left(\partial_\alpha\mathrm{m}^\gamma\partial_\beta\mathrm{m}^\gamma+\partial_\alpha\mathrm{m}_n\partial_\beta\mathrm{m}_n\right)\dd S,\\
        &\tilde{\mathcal{J}}^{(1)}_\alpha=\left(\tilde{\mathcal{G}}_{\alpha\beta}+\mathcal{G}^u_{\alpha\beta}\right)u^\beta,\qquad\tilde{\mathcal{J}}^{(2)}_\alpha=-(\tilde{\mathcal{D}}_{\alpha\beta}+\mathcal{D}^u_{\alpha\beta})u^\beta,\\ &\tilde{\mathcal{G}}_{\alpha\beta}=h\frac{M_s}{\gamma}\int\begin{vmatrix}
            \mathrm{m}^1 & \partial_\alpha\mathrm{m}^1 & \nabla_\beta\mathrm{m}^1 \\
            \mathrm{m}^2 & \partial_\alpha\mathrm{m}^2 & \nabla_\beta\mathrm{m}^2 \\
            \mathrm{m}_n & \partial_\alpha\mathrm{m}_n & \nabla_\beta\mathrm{m}_n
        \end{vmatrix}\dd S,\qquad \mathcal{G}_{\alpha\beta}^u=h\frac{M_s}{\gamma}\int\begin{vmatrix}
            \mathrm{m}^1 & \partial_\alpha\mathrm{m}^1 & \Omega_\beta \mathrm{m}^2 -h_{1\beta}\mathrm{m}_n \\
            \mathrm{m}^2 & \partial_\alpha\mathrm{m}^2 & -\Omega_\beta \mathrm{m}^1 -h_{2\beta}\mathrm{m}_n \\
            \mathrm{m}_n & \partial_\alpha\mathrm{m}_n & h_{\beta\gamma}\mathrm{m}^\gamma
        \end{vmatrix}\dd S,\\
        &\tilde{\mathcal{D}}_{\alpha\beta}=h\frac{M_s}{\gamma}\int\left(\partial_\alpha\mathrm{m}^\gamma\nabla_\beta\mathrm{m}^\gamma+\partial_\alpha\mathrm{m}_n\nabla_\beta\mathrm{m}_n\right)\dd S,\quad \mathcal{D}^u_{\alpha\beta}=h\frac{M_s}{\gamma}\int\left[\partial_\alpha\mathrm{m}^\gamma(\Omega_\beta\epsilon_{\gamma\delta}\mathrm{m}^\delta-h_{\beta\gamma}\mathrm{m}_n)+\partial_\alpha \mathrm{m}_n h_{\beta\gamma}\mathrm{m}^\gamma
        \right]\dd S,
    \end{split}
\end{equation}
where we assumed that $u^\alpha$ does not depend on $\xi^1$ and $\xi^2$. The substitution of \eqref{eq:coefs-Thiele} into \eqref{eq:Thiele-gen} results in the generalized Thiele equation 
\begin{equation}\label{eq:Thiele}
    \mathcal{G}_{ab}\dot{X}^b-\tilde{\mathcal{G}}_{ab}u^b=\frac{\partial\mathcal{H}}{\partial X^a}+\alpha \mathcal{D}_{ab}\dot{X}^b-\beta \tilde{\mathcal{D}}_{ab} u^b+(\mathcal{G}^u_{ab}-\beta \mathcal{D}^u_{ab})u^b.
\end{equation}

%Deriving \eqref{eq:coefs-Thiele} we utilized the following relations for vectors $\vec{a}=\vec{g}_\alpha a^\alpha+\vec{n}a_n$, $\vec{b}=\vec{g}_\alpha b^\alpha+\vec{n}b_n$, and $\vec{c}=\vec{g}_\alpha c^\alpha+\vec{n}c_n$:
%\begin{equation}
%    \vec{a}\cdot[\vec{b}\times\vec{c}]=\sqrt{|g|}\begin{vmatrix}
%        a^1 & a^2 & a_n\\
%        b^1 & b^2 & b_n\\
%        c^1 & c^2 & c_n
%    \end{vmatrix}\qquad \vec{a}\cdot\vec{b}=g_{\alpha\beta}a^\alpha b^\beta+a_nb_n.
%\end{equation}

\section{Geodesic polar coordinates approximation}

Since skyrmions in flat films exhibit rotational symmetry, it is convenient to describe them in polar coordinates $(\rho,\chi)$. To compute the skyrmion energy for a curvilinear film, we can use geodesic polar coordinates (GPC), as follows.

%\subsection{Geodesic polar coordinates}

Let $O$ be the center of some geodesic polar reference frame, and let $\vec{\gamma}_\chi(\rho)$ be a geodesic that goes from $O$ to a point in some direction (e.g., one of the principal directions) describing an angle $\chi$, as shown in Fig. S1 . Here, $\rho>0$ is the natural parameter (arc length) of the geodesic starting at $O$. In this frame, a point $P$ in $\vec{\gamma}_\chi(\rho)$ in the vicinity of $O$ is represented by the geodesic polar coordinates $(\rho,\chi)$, as shown in Fig.~\ref{fig:GPC}, with $\rho$ the geodesic distance between $O$ and $P$. 
\begin{figure}
    \centering
    \includegraphics[width=\linewidth]{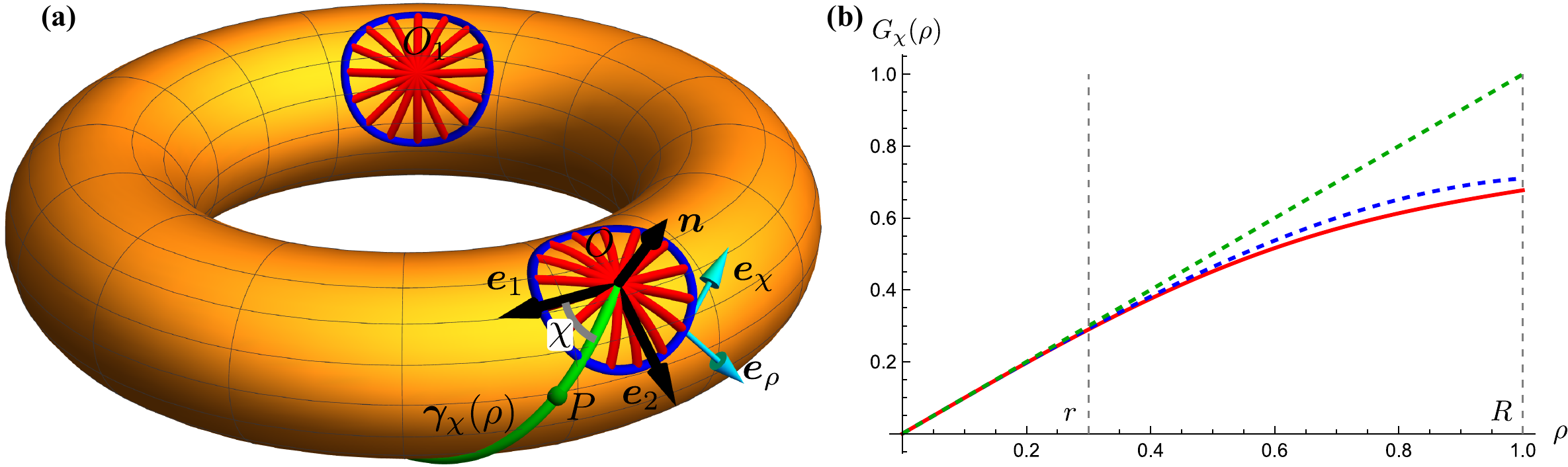}
    \caption{(a) -- Illustration of the geodesic polar coordinates. Red lines depict two geodesics of equal length that  start at the centers $O$ and $O_1$ describing different trajectories. The corresponding geodesic circles are shown in blue. A geodesic $\vec{\gamma}_\chi(\rho)$ runs out of the center $O$ describing an angle $\chi$ relative to the basis vector $\vec{e}_1$. A point $P$ lying on $\vec{\gamma}_\chi(\rho)$ is described by the geodesic polar coordinates $(\rho,\chi)$, where $\rho$ is the geodesic arc length between $P$ and $O$. (b) -- The comparison of the metric function $G_{\pi/4}(\rho)$ (red solid line) computed along geodesic $\vec{\gamma}_{\pi/4}$ to the approximations $G_{\pi/4}(\rho)\approx\rho-\frac{\rho^3}{3!}\mathcal{K}(0)$ (blue dashed line) and $G_{\pi/4}(\rho)\approx\rho$ (green dashed line). The geodesic $\vec{\gamma}_{\pi/4}$ is shown by the green line on the panel (a), it starts from the center $O$ with coordinates $(\xi^1=0,\xi^2=0.2)$. The minor $r$ and major $R$ torus radii are shown by the vertical dashed lines to indicate the typical length scales. }
    \label{fig:GPC}
\end{figure}

Locus of points around $O$ with $\rho=\text{const}$ is the geodesic circle, as depicted Fig.~\ref{fig:GPC}.

According to Gauss's Lemma, the GPC possesses a diagonal metric~\cite{Carmo16,Tristan21}
\begin{equation}\label{eq:CPC-gab}
    [\mathfrak{g}_{\alpha\beta}]=\begin{bmatrix}
        1 & 0 \\
        0 & G^2_\chi(\rho)
    \end{bmatrix}.
\end{equation}
For a given $\chi$, the function $G_\chi(\rho)$ is determined by the ordinary differential equation (Jacobi equation)
\begin{equation}\label{eq:Jacobi}   \partial_{\rho\rho}G_\chi+\mathcal{K}(\vec{\gamma}_\chi(\rho))G_\chi=0,\qquad G_\chi(0)=0,\quad \partial_\rho G_\chi(0)=1 \, ,
\end{equation}
where $\mathcal{K}(\vec{\gamma}_\chi(\rho))$ is the Gaussian curvature along the geodesic $\vec{\gamma}_\chi(\rho)$. In the small vicinity of $O$, the function $G_\chi(\rho)$ can be approximated as follows
\begin{equation}\label{eq:G-approx}
   G_\chi(\rho)\approx\rho-\frac{\rho^3}{3!}\mathcal{K}(0)\approx\rho,
\end{equation}
where $\mathcal{K}(0)$ is the Gaussian curvature at the center $O$. The comparison between the approximations \eqref{eq:G-approx} and an exact solution is illustrated in Fig.~\ref{fig:GPC}(b). As can be seen in the figure, the approximation $G_\chi(\rho)\approx\rho$ is in excellent agreement with the exact solution for $\rho\ll r$.

With the help of the metric \eqref{eq:CPC-gab} we compute the Christoffel symbols of GPC:
\begin{equation}\label{eq:GPC-Gamma}
    \Gamma^\chi_{\rho\chi}=\Gamma^\chi_{\chi\rho}=\frac{\partial_\rho G_{\chi}(\rho)}{ G_{\chi}(\rho)},\qquad \Gamma^\rho_{\chi\chi}=-G_{\chi}(\rho)\partial_\rho G_{\chi}(\rho),\qquad \Gamma^\chi_{\chi\chi}=\frac{\partial_\chi G_{\chi}(\rho)}{ G_{\chi}(\rho)} \, ,
\end{equation}
and the others are zero. The vector of the spin connection is
\begin{equation}
    \vec{\Omega}=\Gamma^\rho_{\rho\chi}\frac{\vec{e}_\rho}{\sqrt{\mathfrak{g}_{\chi\chi}}}-\Gamma^\chi_{\rho\chi}\frac{\vec{e}_\chi}{\sqrt{\mathfrak{g}_{\rho\rho}}}=-\frac{\partial_\rho G_{\chi}(\rho)}{ G_{\chi}(\rho)}\vec{e}_\chi=-\frac{\vec{e}_\chi}{\rho}+\mathcal{O}(\mathcal{K}) \, .
\end{equation}

In terms of the curvilinear coordinates $(\xi^1,\xi^2)$ defined on a surface $\vec{\sigma}=\vec{\sigma}(\xi^1,\xi^2)$, the geodesic $\vec{\gamma}_\chi(\rho)=\vec{\sigma}(\xi^1(\rho),\xi^2(\rho))$ with $\rho>0$, which goes from the point $(\xi^1,\xi^2)=\left(X^1,X^2\right)$ at the angle $\chi$ relative to the direction $\vec{g}_1=\partial_1\vec{\sigma}$ is determined by the following set of equations
\begin{equation}\label{eq:geodesic}
    \begin{split}
        &\ddot{\xi}^\alpha+\Gamma^\alpha_{\beta\gamma}(\xi^1,\xi^2)\dot{\xi}^\beta\dot{\xi}^\gamma=0,\qquad\xi^\alpha(0)=X^\alpha,\quad\dot\xi^\alpha(0)=\tau^\alpha(\chi)\, .
    \end{split}
\end{equation}
In this Eq. a dot denotes the derivative with respect to the arc length $\rho$, and $\vec{\tau}(\chi)$ is the unit vector at $(X^1,X^2)$ which makes an angle $\chi$ with $\vec{g}_1$. For the orthogonal basis $(g_{12}=0)$
\begin{equation}\label{eq:tau}
    \tau^1(\chi)=\frac{\cos\chi}{\sqrt{g_{11}(X^1,X^2)}},\qquad\tau^2(\chi)=\frac{\sin\chi}{\sqrt{g_{22}(X^1,X^2)}}.
\end{equation}

In the vicinity of the point $(X^1,X^2)$, the solution of Eq.~\eqref{eq:geodesic} can be presented in terms of a Taylor series
\begin{equation}\label{eq:xi-approx}
    \xi^\alpha(\rho)\approx\xi^\alpha(0)+\dot{\xi}^\alpha(0)\rho+\ddot{\xi}^\alpha(0)\frac{\rho^2}{2}=X^\alpha+\tau^\alpha(\chi)\rho-\omega^\alpha(\chi)\frac{\rho^2}{2} \, ,
\end{equation}
where $\omega^\alpha(\chi)=\Gamma^\alpha_{\beta\gamma}(X^1,X^2)\tau^\beta(\chi)\tau^\gamma(\chi)$. 
Using \eqref{eq:xi-approx}, we derive
the Jacobian of the transition $\{\xi^1,\xi^2\}\to\{\rho,\chi\}$:
\begin{equation}\label{eq:J}
    J^\alpha_{\alpha'}=\begin{bmatrix}
        \frac{\partial\xi^1}{\partial\rho} &\frac{\partial\xi^1}{\partial\chi} \\
        \frac{\partial\xi^2}{\partial\rho} &\frac{\partial\xi^2}{\partial\chi}
    \end{bmatrix}\approx\begin{bmatrix}
        \tau^1(\chi)-\rho \omega^1(\chi) & \rho\partial_\chi\tau^1(\chi)-\frac{\rho^2}{2}\partial_\chi\omega^1(\chi) \\
         \tau^2(\chi)-\rho \omega^2(\chi)& \rho\partial_\chi\tau^2(\chi)-\frac{\rho^2}{2} \partial_\chi\omega^2(\chi)
    \end{bmatrix}.
\end{equation}
With the help of \eqref{eq:J} we can relate components of the magnetization in different reference frames. Let $\{\vec{e}_1,\vec{e}_2\}$ and $\{\vec{e}_\rho,\vec{e}_\chi\}$ be normalized basis vectors in the coordinates $\{\xi^1,\xi^2\}$ and GPC $\{\rho,\chi\}$, respectively, and $\vec{m}=\vec{e}_\alpha m^\alpha+\vec{n}m_n=\vec{e}_\rho m^\rho+\vec{e}_\chi m^\chi+\vec{n}m_n$, then
\begin{equation}\label{eq:m-alpha}
    m^\alpha=\sqrt{g_{\alpha\alpha}(\xi^1,\xi^2)}\left[\frac{\partial\xi^\alpha}{\partial\rho}m^\rho+\frac{\partial\xi^\alpha}{\partial\chi}\frac{m^\chi}{G_\chi(\rho)}\right].
\end{equation}
Using the Taylor expansion in the vicinity of $(X^1,X^2)$, one obtains
\begin{equation}\label{eq:gaa-sqrt}
    \sqrt{g_{\alpha\alpha}(\xi^1,\xi^{2})}\approx\sqrt{g_{\alpha\alpha}(X^1,X^{2})}\left[1+\Gamma^\alpha_{\alpha\beta}(X^1,X^2)(\xi^\beta-X^\beta)\right]\approx \sqrt{g_{\alpha\alpha}(X^1,X^{2})}\left[1+\Gamma^\alpha_{\alpha\beta}(X^1,X^2)\tau^\beta(\chi)\rho\right].
\end{equation}
Here, we used \eqref{eq:xi-approx} in the last step. So, using \eqref{eq:m-alpha} with \eqref{eq:J} and \eqref{eq:gaa-sqrt}, we can express $m^1$ and $m^2$ via components $m^\rho(\rho,\chi)$, $m^\chi(\rho,\chi)$ and GPC $(\rho,\chi)$.

The inverted derivatives can be obtained from \eqref{eq:J} using the relation 
\begin{equation}\label{eq:ders-inv}
    \begin{bmatrix}
        \frac{\partial\rho}{\partial\xi^1} & \frac{\partial\rho}{\partial\xi^2} \\
          \frac{\partial\chi}{\partial\xi^1} & \frac{\partial\chi}{\partial\xi^2}
    \end{bmatrix}=\begin{bmatrix}
        \frac{\partial\xi^1}{\partial\rho} &\frac{\partial\xi^1}{\partial\chi} \\
        \frac{\partial\xi^2}{\partial\rho} &\frac{\partial\xi^2}{\partial\chi}
    \end{bmatrix}^{-1}.
\end{equation}
With \eqref{eq:ders-inv}, one can compute $\partial_\alpha=\partial_\alpha\rho\partial_\rho+\partial_\alpha\chi\partial_\chi$. 

\subsection{Skyrmion Ansatz}

In GPC, whose center $O$ coincides with the skyrmion center, the skyrmion magnetization can be described with a extremly siple ansatz:
\begin{equation}\label{eq:skAnsatz}
    \vec{\mathrm{m}}=\sin\Theta(\rho)\left[\cos\Phi_0\vec{e}_\rho+\sin\Phi_0\vec{e}_\chi\right]+\cos\Theta(\rho){\vec{n}}.
\end{equation}
Here, $\Theta(\rho)$ is the known profile of the planar skyrmion, and $\Phi_0$ is a constant. For a N{\'e}el skyrmion, $\Phi_0=0,\pi$. The unit vectors $\vec{e}_\rho$ and $\vec{e}_\chi$ form the orhonormal basis of the GPC shown in Fig.~\ref{fig:GPC}(a), and $\vec{n}=\vec{e}_\rho\times\vec{e}_\chi$.

\subsection{Exchange energy}

In an arbitrary orthonormal basis $\{\vec{\mathfrak{e}}_1,\vec{\mathfrak{e}}_2,\vec{n}\}$ the magnetization field $\vec{m}$ can be presented using the parameterization of the spherical angles $0\le\theta\le\pi$ and $0\le\phi<2\pi$:
\begin{equation}\label{eq:ang-curv}
    \vec{m}=\sin\theta\vec{\mu}(\phi)+\cos\theta\vec{n},\qquad\vec{\mu}(\phi)=\cos\phi\vec{\mathfrak{e}}_1+\sin\phi\vec{\mathfrak{e}}_2.
\end{equation}
Here, $\vec{\mu}$ is the normalized tangential magnetization. In terms of $\theta$ and $\phi$, the exchange energy density $\mathscr{E}_{ex}=-A\vec{m}\cdot\Delta\vec{m}$ can be written as follows~\cite{Gaididei-PRL}
\begin{equation}\label{eq:Eex}
    \mathscr{E}_{ex}=A_{ex}\left\{[\vec{\nabla}\theta-\vec{\Gamma}]^2+\left[\sin\theta(\vec{\nabla}\phi-\vec{\Omega})-\cos\theta\,\partial_\phi\vec{\Gamma}\right]^2\right\} \, ,
\end{equation}
where $\vec{\Gamma}=\vec{\mathfrak{e}}_\alpha h_{\alpha\beta}\mu_\beta$, where $h_{\alpha\beta}$ is the shape operator (aka Weingarten map). In an orthonormalized basis, it has the following form
\begin{equation}\label{eq:Weingarten}
    [h_{\alpha\beta}]=\begin{bmatrix}
        \frac{b_{11}}{g_{11}} & \frac{b_{12}}{\sqrt{g_{11}g_{22}}} \\
        \frac{b_{21}}{\sqrt{g_{11}g_{22}}} & \frac{b_{22}}{g_{22}}
    \end{bmatrix}=\begin{bmatrix}
        \kappa_1\cos^2\tilde{\chi}+\kappa_2\sin^2\tilde{\chi} & (\kappa_1-\kappa_2)\sin\tilde{\chi}\cos\tilde{\chi}\\
        (\kappa_1-\kappa_2)\sin\tilde{\chi}\cos\tilde{\chi} & \kappa_1\sin^2\tilde{\chi}+\kappa_2\cos^2\tilde{\chi} 
    \end{bmatrix},
\end{equation}
where $\kappa_\alpha$ are the principal curvatures determined as the eigenvalues of $[h_{\alpha\beta}]$, and $\tilde{\chi}$ is the angle of rotation of the tangential basis $\{\vec{\mathfrak{e}}_1,\vec{\mathfrak{e}}_2\}$ relative to the principal basis $\{\vec{\nu}_1,\vec{\nu}_2\}$, {\it i.e.}
\begin{equation}
\begin{bmatrix}
    \vec{\mathfrak{e}}_1 \\ \vec{\mathfrak{e}}_2
\end{bmatrix}=\begin{bmatrix}
    \cos\tilde{\chi} & \sin\tilde{\chi} \\
    -\sin\tilde{\chi} & \cos\tilde{\chi}
\end{bmatrix}\begin{bmatrix}
    \vec{\nu}_1 \\ \vec{\nu}_2
\end{bmatrix}.
\end{equation}
In this Eq. $\vec{\nu}_\alpha$ are the normalized eigenvectors of $[h_{\alpha\beta}]$ such that $\vec{\nu}_1\times\vec{\nu}_2=\vec{n}$. Note that the last part of \eqref{eq:Weingarten} is the general representation of a symmetric 2nd-rank tensor in a orthorormalized basis rotated relative to the tensor's eigen-directions.

For the skyrmion Ansatz~\eqref{eq:skAnsatz} defined in GPC one has $\vec{\mathfrak{e}}_1=\vec{e}_\rho$, $\vec{\mathfrak{e}}_2=\vec{e}_\chi$, $\vec{\nabla}\theta=\Theta'(\rho)\vec{e}_\rho$ and $\vec{\nabla}\phi=\vec{0}$. Neglecting terms of the order of the Gaussian curvature, we also obtain $\vec{\Omega}\approx-\vec{e}_\chi/\rho$, $\tilde{\chi}\approx\chi$, and $\sqrt{\det[\mathfrak{g}_{\alpha\beta}]}\approx\rho$~\footnote{Note that neglecting the Gaussian curvature in the Jacobi equation~\eqref{eq:Jacobi}, we obtain $G_\chi(\rho)=\rho$, which corresponds to the planar polar coordinates.}. 

Finally, we write \eqref{eq:Eex} in the form
\begin{equation}\label{eq:Eex-den}
    \begin{split}\mathscr{E}_{ex}=A_{ex}\biggl\{\Theta'^2+\frac{\sin^2\Theta}{\rho^2}&-2\cos\Phi_0\left[\Theta'(\kappa_1\cos^2\chi+\kappa_2\sin^2\chi)+\frac{\sin\Theta\cos\Theta}{\rho}(\kappa_1\sin^2\chi+\kappa_2\cos^2\chi)\right]\\
    &-\sin\Phi_0\sin2\chi(\kappa_1-\kappa_2)\left(\Theta'-\frac{\sin\Theta\cos\Theta}{\rho}\right)+\mathcal{O}(\kappa_1^2,\kappa_2^2,\kappa_1\kappa_2)\biggr\}
    \end{split}
\end{equation}
Note that in order to keep terms quadratic in principal curvatures, one needs to take into account the deviation of the metric $\mathfrak{g}_{\alpha\beta}$ from the metric of the planar polar coordinates. The principal curvatures $\kappa_\alpha$ are coordinate-dependent. Using the exponential localization of functions $\Theta'$ and $\sin\Theta$, we expand $\kappa_\alpha$ in a Taylor series in the vicinity of the skyrmion center. In this series, however, we are allowed to keep only leading terms $\kappa_\alpha (0)$, which are the values of $\kappa_\alpha$ in the skyrmion center, because the derivatives $\partial_\beta\kappa_\alpha$ are generally of the order $\mathcal{O}(\kappa_1^2,\kappa_2^2,\kappa_1\kappa_2)$. 

Taking into account the exponential localization of the skyrmion profile, in the limit $R_s\kappa_\alpha\ll1$ with $R_s$ being the skyrmion radius, we compute the total exchange energy as follows
\begin{equation}
    E_{ex}\approx h\int\limits_0^{2\pi}\dd\chi\int\limits_0^\infty\dd\rho G_{\chi}(\rho)\mathscr{E}_{ex}=h\int\limits_0^{2\pi}\dd\chi\int\limits_0^\infty\dd\rho \rho\mathscr{E}_{ex}+\mathcal{O}(\mathcal{K}) \, ,
\end{equation}
where $h$ denotes the thickness of the film. Now, using \eqref{eq:Eex-den} we obtain
\begin{equation}
    E_{ex}=E^0_{ex}-8\pi A_{ex}h\cos\Phi_0\,\mathscr{H}(X^1,X^2)\mathcal{C}_1+\mathcal{O}(\kappa_1^2,\kappa_2^2,\kappa_1\kappa_2),
\end{equation}
where $\mathscr{H}(X^1,X^2)=\kappa_1(X^1,X^2)+\kappa_2(X^1,X^2)$ is the mean curvature at the center of the skyrmion. Here $E^0_{ex}=8\pi A_{ex}h \mathcal{C}_0$ is the total exchange energy of the planar skyrmion, and constants $\mathcal{C}_n$ are 
\begin{equation}
    \mathcal{C}_0=\frac{1}{4}\int\limits_0^{\infty}\dd\rho \rho\left[\Theta'^2+\frac{\sin^2\Theta}{\rho^2}\right],\qquad\mathcal{C}_1=\frac14\int\limits_0^\infty\dd\rho\rho\left[\Theta'+\frac{\sin\Theta\cos\Theta}{\rho}\right].
\end{equation}
For the N{\'e}el skyrmion, $\cos\Phi_0=\pm1$. The developed approach cannot be applied to Bloch skyrmions with $\Phi_0=\pm\frac{\pi}{2}$, because the curvature corrections to the total energy are of the order $\mathcal{O}(\kappa_1^2,\kappa_2^2,\kappa_1\kappa_2)$.

\subsection{DMI energy}
We now consider the interfacial DMI whose energy density is $\mathscr{E}_{\textsc{dm}}=D_{\textsc{dm}}[m_n(\vec{\nabla}\cdot\vec{m})-(\vec{m}\cdot\vec{\nabla})m_n]$. In the angular parameterization~\eqref{eq:ang-curv}, it is defined by the following expression~\cite{Kravchuk-PRL}
\begin{equation}\label{eq:E_DMI_curv}
    \mathscr{E}_{\textsc{dm}}=D_{\textsc{dm}}\left\{2\sin^2\theta(\vec{\mu}\cdot\vec{\nabla}\theta)-\mathscr{H}\cos^2\theta+\vec{\nabla}\cdot(\vec{\mu}\sin\theta\cos\theta)\right\}.
    \end{equation}
Since $\sin\theta$ vanishes at large distances from the center of the skyrmion, the last divergent summand in \eqref{eq:E_DMI_curv} does not contribute to the total energy. Using the skyrmion Ansatz for GPC~\eqref{eq:skAnsatz} and making the same assumptions as in the previous section, we obtain~\cite{Korniienko2020}
\begin{equation}
    E_{\textsc{dm}}=E^0_{\textsc{dm}}+8\pi Dh\mathscr{H}(X^1,X^2)\mathcal{C}_2-D_{\textsc{dm}}h\int \mathscr{H}\dd S+\mathcal{O}(\kappa_1^2,\kappa_2^2,\kappa_1\kappa_2),
\end{equation}
where $E^0_{\textsc{dm}}=4\pi D_{\textsc{dm}}h\cos\Phi_0\int_0^\infty\dd\rho\rho\Theta'\sin^2\Theta$ is DMI energy of the planar skyrmion and
\begin{equation}
    \mathcal{C}_2=\frac14\int\limits_0^\infty\dd\rho\rho\sin^2\Theta.
\end{equation}

\subsection{Total energy}
The anisotropy energy is trivial $E_{an}=Kh\int(1-m_n^2)\dd S=8\pi Kh\mathcal{C}_2+\mathcal{O}(\kappa_1^2\kappa_2^2,\kappa_1\kappa_2)$ and up to the terms quadratic in the principal curvatures does not depend on the skyrmion position. The total energy of a N{\'e}el skyrmion with $\Phi_0=0$ is
\begin{equation}\label{eq:Htot}
    \mathcal{H}=E_0-8\pi h\mathscr{H}(X^1,X^2)(A_{ex}\mathcal{C}_1-D_{\textsc{dm}}\mathcal{C}_2)+\mathcal{O}(\kappa_1^2\kappa_2^2,\kappa_1\kappa_2),
\end{equation}
where the constant $E_0$ is independent on the skyrmion coordinates $(X^1,X^2)$. 

\section{The torus parameterization}
We parameterize our toroidal surface as follows:
\begin{equation}\label{eq:torus}
    \vec{\sigma}(\xi^1,\xi^2)=\vec{e}_x\sin\frac{\xi^1}{R}\left(R+r\sin\frac{\xi^2}{r}\right)+\vec{e}_y\cos\frac{\xi^1}{R}\left(R+r\sin\frac{\xi^2}{r}\right)+\vec{e}_z r\cos\frac{\xi^2}{r},\qquad0\le\xi^1<2\pi R,\quad0\le\xi^2<2\pi r,
\end{equation}
where $R$ and $r$ are the major and minor torus radii, respectively. Parameterization \eqref{eq:torus} induces metric
\begin{equation}
    [g_{\alpha\beta}]=\begin{bmatrix}
        \frac{\left(R+r\sin\frac{\xi^2}{r}\right)^2}{R^2} & 0 \\
        0 &1
    \end{bmatrix}.
\end{equation}
The nonzero Christoffel symbols are
\begin{equation}\label{eq:Gamma-torus}
    \Gamma^1_{12}=\Gamma^1_{21}=\frac{\cos\frac{\xi^2}{r}}{R+r\sin\frac{\xi^2}{r}},\qquad\Gamma^2_{11}=-\frac{\cos\frac{\xi^2}{r}\left(R+r\sin\frac{\xi^2}{r}\right)}{R^2}.
\end{equation}
Vectors $\vec{e}_\alpha=\vec{g}_\alpha/\sqrt{g_{\alpha\alpha}}$ with $\vec{g}_\alpha=\partial_\alpha\vec{\sigma}$, and $\vec{n}=\vec{e}_1\times\vec{e}_2$ compose the orthonomalized basis whose explicit form is
\begin{equation}
    \begin{split}
&\vec{e}_1=\vec{e}_x\cos\frac{\xi^1}{R}-\vec{e}_y\sin\frac{\xi^1}{R},\qquad\vec{e}_2=\vec{e}_x\sin\frac{\xi^1}{R}\cos\frac{\xi^2}{r}+\vec{e}_y\cos\frac{\xi^1}{R}\cos\frac{\xi^2}{r}-\vec{e}_z\sin\frac{\xi^2}{r},\\
&\vec{n}=\vec{e}_x\sin\frac{\xi^1}{R}\sin\frac{\xi^2}{r}+\vec{e}_y\cos\frac{\xi^1}{R}\sin\frac{\xi^2}{r}+\vec{e}_z\cos\frac{\xi^2}{r}.
    \end{split}
\end{equation}

The vector of the spin connection is
\begin{equation}
    \vec{\Omega}=\vec{e}_\alpha(\vec{e}_1\cdot\nabla_\alpha\vec{e}_2)=\Gamma^1_{12}\frac{\vec{e}_1}{\sqrt{g_{22}}}-\Gamma^2_{12}\frac{\vec{e}_2}{\sqrt{g_{11}}}=\frac{\cos\frac{\xi^2}{r}}{R+r\sin\frac{\xi^2}{r}}\vec{e}_1,
\end{equation}

The shape operator
\begin{equation}
    [h_{\alpha\beta}]=[\vec{n}\cdot\nabla_\alpha\vec{e}_\beta]=\begin{bmatrix}
        \kappa_1 & 0\\
        0 &\kappa_2
    \end{bmatrix},\qquad     \kappa_1=-\frac{\sin\frac{\xi^2}{r}}{R+r\sin\frac{\xi^2}{r}},\quad\kappa_2=-\frac{1}{r}
\end{equation}
is diagonal. This means that $\kappa_1$ and $\kappa_2$ are the principal curvatures, and directions of $\vec{e}_1$ and $\vec{e}_2$ coincide with the corresponding principal directions. The Gaussian and mean curvatures are determined as $\mathcal{K}=\kappa_1\kappa_2$ and $\mathscr{H}=\kappa_1+\kappa_2$, respectively.

Using \eqref{eq:m-alpha}, \eqref{eq:gaa-sqrt}, and \eqref{eq:J}, we write the magnetization components $m^1$ and $m^2$ defined in the normalized basis $\{\vec{e}_1,\vec{e}_2\}$ of the coordinates $(\xi^1,\xi2)$ via components $m^\rho$ and $m^\chi$ defined in the normalized basis $\{\vec{e}_\rho,\vec{e}_\chi\}$ of GPC:
\begin{equation}\label{eq:m12}
\begin{split}
    &m^1\approx m^\rho\cos\chi-m^\chi\sin\chi-\frac{\Gamma^1_{12}(X^1,X^2)}{\sqrt{g_{22}(X^1,X^2)}}\rho\cos\chi(m^\rho\sin\chi+m^\chi\cos\chi),\\
    &m^2\approx m^\rho\sin\chi+m^\chi\cos\chi-\frac{\sqrt{g_{22}(X^1,X^2)}\Gamma^2_{11}(X^1,X^2)}{g_{11}(X^1,X^2)}\rho\cos\chi(m^\rho\cos\chi-m^\chi\sin\chi).
    \end{split}
\end{equation}
Here, we neglected terms of second order in the principal curvatures and higher. We also took into account that there are only two nonzero Christoffel symbols \eqref{eq:Gamma-torus} for the toroidal surface \eqref{eq:torus} and that these Christoffel symbols are of the first order in the principal curvatures. Using \eqref{eq:ders-inv} we compute the derivatives
\begin{equation}\label{eq:ders}
\begin{split}
        \partial_1=&\left(\sqrt{g_{11}(X^1,X^2)}\cos\chi+\frac{g_{11}(X^1,X^2)\Gamma^1_{12}(X^1,X^2)+g_{22}(X^1,X^2)\Gamma^2_{11}(X^1,X^2)}{\sqrt{g_{11}(X^1,X^2)g_{22}(X^1,X^2)}}\rho\cos\chi\sin\chi\right)\partial_\rho\\
        +&\left(-\frac{\sqrt{g_{11}(X^1,X^2)}\sin\chi}{\rho}+\frac{g_{22}(X^1,X^2)\Gamma^2_{11}(X^1,X^2)\cos^2\chi-g_{11}(X^1,X^2)\Gamma^1_{12}(X^1,X^2)\sin^2\chi}{\sqrt{g_{11}(X^1,X^2)g_{22}(X^1,X^2)}}\right)\partial_\chi,\\
        \partial_2=&\left(\sqrt{g_{22}(X^1,X^2)}\sin\chi+\Gamma^1_{12}(X^1,X^2)\rho\cos^2\chi\right)\partial_\rho+\left(\frac{\sqrt{g_{22}(X^1,X^2)}\cos\chi}{\rho}-\Gamma^1_{12}(X^1,X^2)\sin\chi\cos\chi\right)\partial_\chi
\end{split}
\end{equation}
Now, with the help of \eqref{eq:ders} and \eqref{eq:m12}, taking into account that $m^\rho=\sin\Theta(\rho)\cos\Phi_0$, $m^\chi=\sin\Theta(\rho)\sin\Phi_0$, and $m_n=\cos\Theta(\rho)$ for a skyrmion, one can show that in the linear order in the Christoffel symbols
\begin{equation}\label{eq:det-approx}
    \begin{vmatrix}
            \mathrm{m}^1 & \partial_\alpha\mathrm{m}^1 & \nabla_\beta\mathrm{m}^1 \\
            \mathrm{m}^2 & \partial_\alpha\mathrm{m}^2 & \nabla_\beta\mathrm{m}^2 \\
            \mathrm{m}_n & \partial_\alpha\mathrm{m}_n & \nabla_\beta\mathrm{m}_n
        \end{vmatrix}=\frac{1}{\sqrt{g_{\beta\beta}(X^1,X^2)}}\begin{vmatrix}
            \mathrm{m}^1 & \partial_\alpha\mathrm{m}^1 & \partial_\beta\mathrm{m}^1 \\
            \mathrm{m}^2 & \partial_\alpha\mathrm{m}^2 & \partial_\beta\mathrm{m}^2 \\
            \mathrm{m}_n & \partial_\alpha\mathrm{m}_n & \partial_\beta\mathrm{m}_n
        \end{vmatrix}-\frac{\sqrt{g_{11}(X^1,X^2)g_{22}(X^1,X^2)}}{\sqrt{g_{\beta\beta}(X^1,X^2)}}\Gamma^\beta_{\beta\gamma}(X^2,X^2)\tau^\gamma\epsilon_{\alpha\beta}\sin\Theta\Theta'.
\end{equation}
In this equation we used the relation 
\begin{equation}
    \frac{1}{\sqrt{g_{\alpha\alpha}(\xi^1,\xi^2)}}\approx\frac{1}{\sqrt{g_{\alpha\alpha}(X^1,X^2)}}\left[1-\Gamma^\alpha_{\alpha\beta}(X^1,X^2)\tau^\beta\rho\right].
\end{equation}
The last summand in \eqref{eq:det-approx} is averaged out during the integration over $\chi$ because of the form of $\tau^\gamma$, see \eqref{eq:tau}. Note that in the linear order in the principal curvatures, $\dd S=G_\chi(\rho)\dd\rho\dd\chi\approx\rho\dd\rho\dd\chi$. Thus, in the linear order in the principal curvatures we obtain $\tilde{\mathcal{G}}_{\alpha\beta}\approx \mathcal{G}_{\alpha\beta}/\sqrt{g_{\beta\beta}(X^1,X^2)}$. It is important to note that the relation \eqref{eq:det-approx} is valid for any surface, not necessarily toroidal.  In the same way, one can demonstrate that $\tilde{\mathcal{D}}_{\alpha\beta}\approx \mathcal{D}_{\alpha\beta}/\sqrt{g_{\beta\beta}(X^1,X^2)}$ for any surface. Using these relations, we write the Thiele equations \eqref{eq:Thiele} in the form
\begin{equation}\label{eq:Thiele-main}
   \boxed{ G_{ab}(V^b-u^b)=-F_a+D_{ab}(\alpha V^b-\beta u^b)+(G^u_{ab}-\beta D^u_{ab})u^b,}
\end{equation}
where $V^\alpha=\sqrt{g_{\alpha\alpha}(X^1,X^2)}\dot{X}^\alpha$ are components of the skyrmion velocity during its motions along the surface, 
\begin{equation}
    F_\alpha=-\frac{1}{\sqrt{g_{\alpha\alpha}(X^1,X^2)}}\frac{\partial\mathcal{H}}{\partial X^\alpha}
\end{equation}
is the force acting on skyrmion, and we defined the following tensors $G_{\alpha\beta}=\mathcal{G}_{\alpha\beta}/\sqrt{g_{11}(X^1,X^2)g_{22}(X^1,X^2)}$, $D_{\alpha\beta}=\mathcal{D}_{\alpha\beta}/g_{\alpha\alpha}(X^1,X^2)$, $D^u_{\alpha\beta}=\mathcal{D}^u_{\alpha\beta}/\sqrt{g_{\alpha\alpha}(X^1,X^2)}$, and $G^u_{\alpha\beta}=\mathcal{G}^u_{\alpha\beta}/\sqrt{g_{\alpha\alpha}(X^1,X^2)}$.

Using the skyrmion Ansatz \eqref{eq:skAnsatz} together with formulas \eqref{eq:m12} and \eqref{eq:ders} generalized for an arbitrary surface, 8{\it i.e.}, all 6 Christoffel symbols $\Gamma^\alpha_{\beta\gamma}$ are taken into account), in the linear order in the principal curvatures, we obtain 
\begin{equation}
    [G_{\alpha\beta}]=4\pi N_{\mathrm{top}}\frac{hM_s}{\gamma}\begin{bmatrix}
        0 & 1 \\
        -1 & 0
    \end{bmatrix},\qquad      [D_{\alpha\beta}]=4\pi\frac{hM_s}{\gamma}\mathcal{C}_0\begin{bmatrix}
        1 & 0 \\
        0 & 1
    \end{bmatrix}.
\end{equation}
Here $N_{\mathrm{top}}=\frac12\left[\cos\Theta(0)-\cos\Theta(\infty)\right]$ is the skyrmion topological charge. And the current-induced corrections are
\begin{equation}
    [G_{\alpha\beta}^u]=\begin{bmatrix}
        0 & 0 \\
        0 & 0
    \end{bmatrix},\qquad [D_{\alpha\beta}^u]=-4\pi\frac{hM_s}{\gamma}\mathcal{C}_1\begin{bmatrix}
        h_{11}\cos\Phi_0+h_{12}\sin\Phi_0 & h_{12}\cos\Phi_0+h_{22}\sin\Phi_0 \\
        h_{12}\cos\Phi_0-h_{11}\sin\Phi_0 & h_{22}\cos\Phi_0-h_{12}\sin\Phi_0 
    \end{bmatrix}.
\end{equation}
%where 
%\begin{equation}
%    \mathcal{C}_3=\frac14\int\limits_0^\infty\left(\Theta'+\frac{\sin\Theta\cos\Theta}{\rho}\right)\dd\rho.
%\end{equation}
Here, $h_{\alpha\beta}=h_{\alpha\beta}(X^1,X^2)$ are values of the shape operator elements in the skyrmion center.  For a N{\'e}el skyrmion, one has $\sin\Phi_0=0$ and we obtain $D^u_{\alpha\beta}=-4\pi\frac{hM_s}{\gamma}\mathcal{C}_1h_{\alpha\beta}$.

Taking into account that $h_{11}=\kappa_1$, $h_{22}=\kappa_2$ and $h_{12}=0$ for the considered toroidal surface, we write Thiele equations~\eqref{eq:Thiele-main} in the following explicit form
\begin{subequations}\label{eq:EQS}
    \begin{align}
        &N_{\mathrm{top}}\left(V^2-u^2\right)=-\mathcal{F}_1+\alpha\mathcal{C}_0V^1-\beta u^1(\mathcal{C}_0-\kappa_1\mathcal{C}_1),\\
        -&N_{\mathrm{top}}\left(V^1-u^1\right)=-\mathcal{F}_2+\alpha\mathcal{C}_0V^2-\beta u^2(\mathcal{C}_0-\kappa_2\mathcal{C}_1)\, ,
    \end{align}
\end{subequations}
where 
\begin{equation}
    \mathcal{F}_\alpha=\frac{1}{4\pi}\frac{\gamma}{hM_s}F_\alpha=\frac{2\gamma}{M_s}\frac{A_{ex}\mathcal{C}_1-D_{\textsc{dm}}\mathcal{C}_2}{\sqrt{g_{\alpha\alpha}(X^1,X^2)}}\frac{\partial \mathscr{H}(X^1,X^2)}{\partial X^\alpha}
\end{equation}
is the normalized curvature-induced force acting on skyrmion. For the considered torus surface,
\begin{equation}
    \mathcal{F}_1=0,\qquad\mathcal{F}_2=-\frac{2\gamma}{M_s}(A_{ex}\mathcal{C}_1-D_{\textsc{dm}}\mathcal{C}_2)\frac{R\cos\frac{X^2}{r}}{r\left(R+r\sin\frac{X^2}{r}\right)^2}.
\end{equation}
In this case, without current and damping, Eqs.~\eqref{eq:EQS} have solution $V^2=0$ and $V^1=\mathcal{F}_2/N_{\mathrm{top}}$, which is equivalent to
\begin{equation}
    X^2=\mathrm{const},\qquad \dot{X}^1=-\frac{1}{N_{\mathrm{top}}}\frac{2\gamma}{M_s}(A_{ex}\mathcal{C}_1-D_{\textsc{dm}}\mathcal{C}_2)\frac{R^2\cos\frac{X^2}{r}}{r\left(R+r\sin\frac{X^2}{r}\right)^3}.
\end{equation}
{\it i.e.} the skyrmion moves with a constant velocity $V^1(X^2)=\mathcal{F}_2(X^2)/N_{\mathrm{top}}$ which depends on the coordinate $X^2$. Skyrmion makes one complete rotation per period $T=2\pi R/\dot{X}^1$.

In the general case, when the current is applied along $\vec{e}_1$, Eqs.~\ref{eq:EQS} can be formulated in the form of the following set of ODE
\begin{equation}\label{eq:ODE}
\begin{split}
    &\dot{X}^1=\frac{1}{\sqrt{g_{11}(X^1,X^2)}}\frac{N_{\mathrm{top}}\mathcal{F}_2(X^1,X^2)+u^1\left\{N_{\mathrm{top}}^2+\alpha\beta\mathcal{C}_0[\mathcal{C}_0-\mathcal{C}_1\kappa_1(X^1,X^2)]\right\}}{N_{\mathrm{top}}^2+\alpha^2\mathcal{C}_0^2},\\
    &\dot{X}^2=\frac{1}{\sqrt{g_{22}(X^1,X^2)}}\frac{\alpha\mathcal{C}_0\mathcal{F}_2(X^1,X^2)+u^1N_{\mathrm{top}}\left\{\mathcal{C}_0(\alpha-\beta)+\beta\mathcal{C}_1\kappa_1(X^1,X^2)\right\}}{N_{\mathrm{top}}^2+\alpha^2\mathcal{C}_0^2}\, ,
    \end{split}
\end{equation}

which should be solved with some initial conditions $X^1(0)=X^1_0$ and $X^2(0)=X^2_0$.

\subsubsection{Dimensionless units}
Introducing the dimensionless coordinates $\tilde{X}^\alpha=X^\alpha/\ell$ and time $\tilde{t}=\omega_0t$, where $\ell=\sqrt{A_{ex}/K}$ and $\omega_0=2\gamma K/M_s$ with $K$ being the easy-axial anisotropy, we reduce the number of the control parameters in \eqref{eq:ODE} to three: the dimensionless DM constant $d=D_{\textsc{dm}}/\sqrt{A_{ex}K}$, and two dimensionless radii $\tilde{r}=r/\ell$, and $\tilde{R}=R/\ell$. Eqs.~\eqref{eq:ODE} obtain the following form 
\begin{equation}\label{eq:ODE-dimless}
\begin{split}
    &\partial_{\tilde{t}}\tilde{X}^1=\frac{1}{\sqrt{g_{11}}}\frac{N_{\mathrm{top}}\tilde{\mathcal{F}}_2+\tilde{u}^1\left\{1+\alpha\beta\mathcal{C}_0[\mathcal{C}_0-\tilde{\mathcal{C}}_1\tilde{\kappa}_1]\right\}}{1+\alpha^2\mathcal{C}_0^2},\\
    &\partial_{\tilde{t}}\tilde{X}^2=\frac{1}{\sqrt{g_{22}}}\frac{\alpha\mathcal{C}_0\tilde{\mathcal{F}}_2+\tilde{u}^1N_{\mathrm{top}}\left\{\mathcal{C}_0(\alpha-\beta)+\beta\tilde{\mathcal{C}}_1\tilde{\kappa}_1\right\}}{1+\alpha^2\mathcal{C}_0^2},
    \end{split}
\end{equation}
where $\tilde{u}^\alpha=u^\alpha/u_0$ with $u_0=\ell\omega_0$, $\tilde{\kappa}_\alpha=\ell\kappa_\alpha$ and 
\begin{equation}
    \tilde{\mathcal{F}}_2=-\tilde{\Xi}\frac{\tilde{R}\cos\frac{\tilde{X}^2}{\tilde{r}}}{\tilde{r}\left(\tilde{R}+\tilde{r}\sin\frac{\tilde{X}^2}{\tilde{r}}\right)^2}\, ,
\end{equation}
with $\tilde{\Xi}=\tilde{\mathcal{C}}_1-d\tilde{\mathcal{C}}_2$ with $\tilde{\mathcal{C}}_n=\mathcal{C}_n/\ell^n$, and we omitted the dependencies on $(\tilde{X}^1,\tilde{X}^2)$ for simplicity. We also took into account that $N^2_{\mathrm{top}}=1$.

For the considered dimensionless units, the profile of a planar skyrmion $\Theta(\tilde{\rho})$ is uniquely determined by the parameter $d$ by means of the differential equation~\cite{Kravchuk18}
\begin{equation}\label{eq:Profile}
    \partial_{\tilde{\rho}\tilde{\rho}}\Theta+\frac{1}{\tilde{\rho}}\partial_{\tilde{\rho}}\Theta-\sin\Theta\cos\Theta\left(1+\frac{1}{\tilde{\rho}^2}\right)+\frac{|d|}{\tilde{\rho}}\sin^2\Theta=0
\end{equation}
supplemented with the boundary conditions $\Theta(0)=\pi$, $\Theta(\infty)=0$. The solution of \eqref{eq:Profile} exists for $|d|<4/\pi$. The examples of profiles $\Theta(\tilde{\rho})$ for different values of $d$ are shown in Fig.~\ref{fig:profile-Rs}(a).
\begin{figure}
    \centering
    \includegraphics[width=0.8\linewidth]{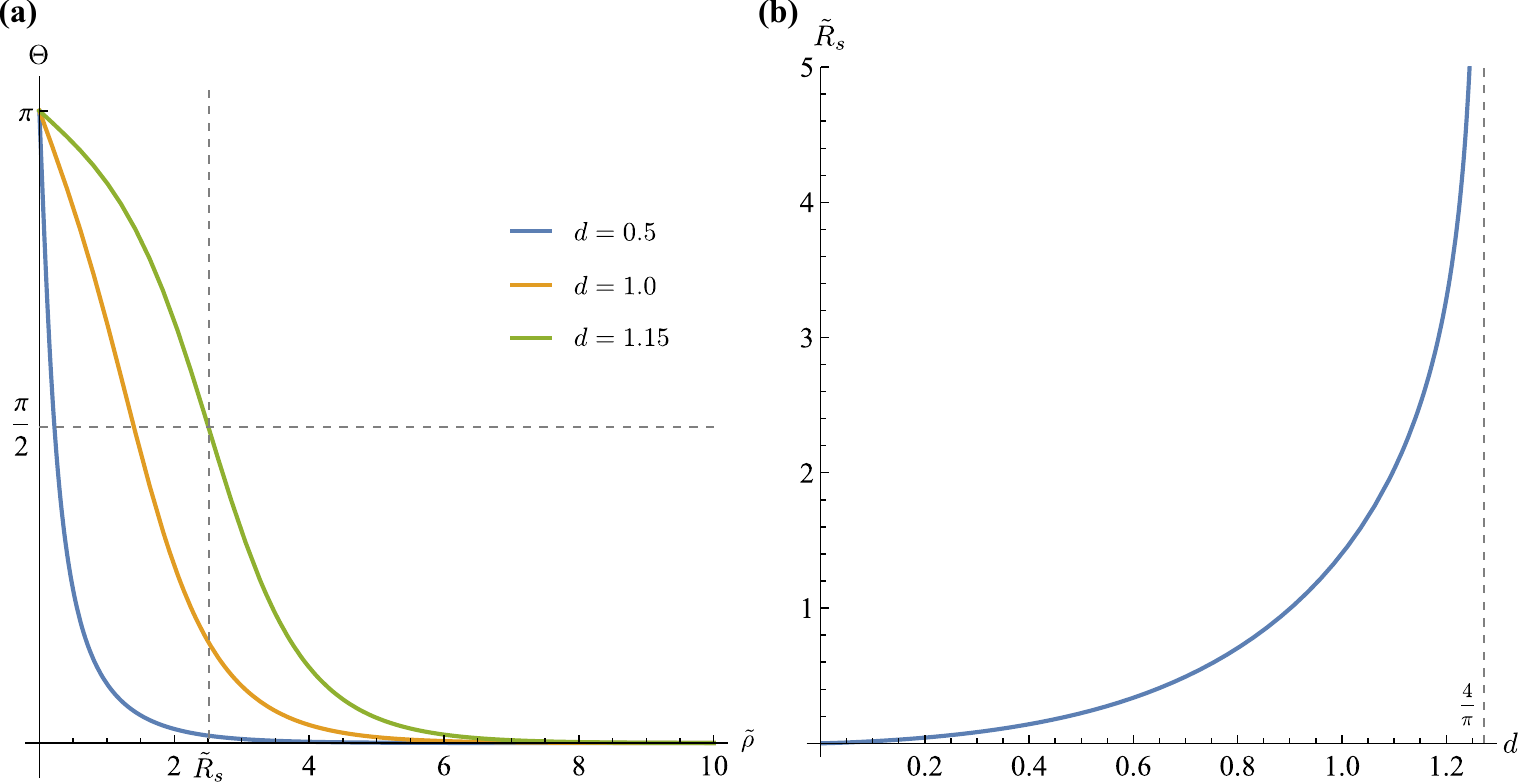}
    \caption{(a) -- profiles of a planar skyrmion obtained as solutions of Eq.~\eqref{eq:Profile} for different values of the dimensionless DMI $d$. (b) -- skyrmion radius as a function of $d$.}
    \label{fig:profile-Rs}
\end{figure}
Having a skyrmion profile, we determine its radius using  $\Theta(\tilde{R}_s)=\pi/2$. The dependence $\tilde{R}_s(d)$ is shown in Fig.~\ref{fig:profile-Rs}(b).

For each given $d$, we obtain the skyrmion profile $\Theta(\tilde{\rho})$ from \eqref{eq:Profile} and then compute coefficients $\tilde{\mathcal{C}}_n$ and $\tilde{\Xi}$, see Fig.~\ref{fig:Cn-Xi}(a).
\begin{figure}
    \centering
    \includegraphics[width=0.8\linewidth]{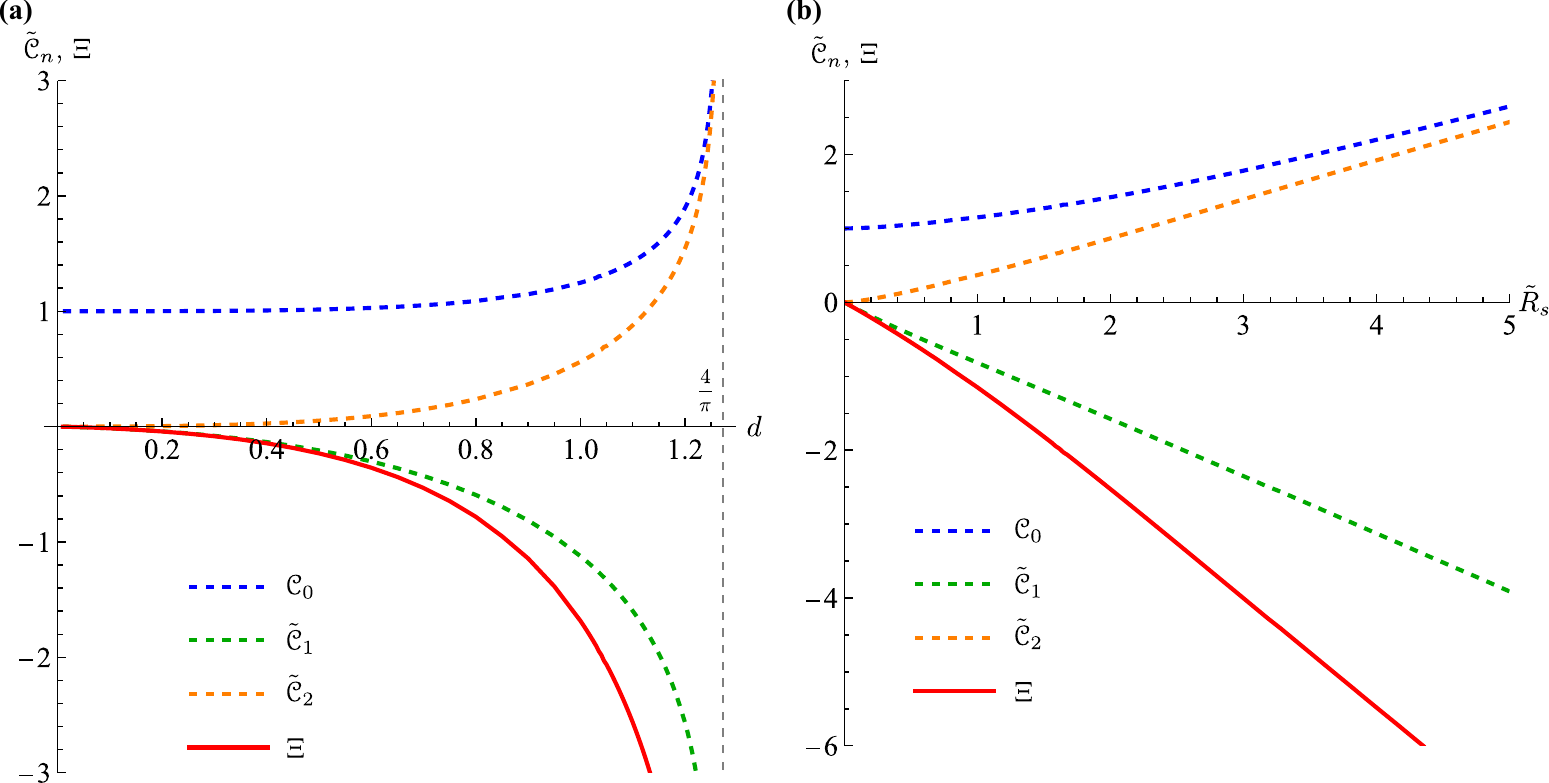}
    \caption{Parameters $\tilde{\mathcal{C}}_n$ and $\tilde{\Xi}$ as functions of the dimensionless DMI $d$ and skyrmion radius $\tilde{R}_s$ are shown in panels (a) and (b), respectively.}
    \label{fig:Cn-Xi}
\end{figure}
using one-to-one correspondence between $d$ and $\tilde{R}_s$ shown in Fig.\ref{fig:profile-Rs}(b), we can alternatively present parameters $\tilde{\mathcal{C}}_n$ and $\tilde{\Xi}$ as functions of the skyrmion radius, see Fig.~\ref{fig:Cn-Xi}(b).

Using \eqref{eq:Htot}, one writes the total skyrmion energy as 
\begin{equation}\label{eq:Htot-Xi}
    \mathcal{H}=E_0-8\pi A h\tilde{\mathscr{H}}\tilde{\Xi}+\mathcal{O}(\kappa_1^2\kappa_2^2,\kappa_1\kappa_2).
\end{equation}
Using \eqref{eq:Htot-Xi}, we plot the dependence of the coordinate-dependent part $\Delta E=\mathcal{H}-E_0$ of the skyrmion energy on the poloidal coordinate in Fig.~2(a,b).%\ref{fig:E-vs-X2}.
% \begin{figure}
%     \centering
%     \includegraphics[width=0.5\linewidth]{E_vs_X2.pdf}
%     \caption{Dependencies of the coordinate-dependent part $\Delta E=\mathcal{H}-E_0$ of the skyrmion energy \eqref{eq:Htot-Xi} on the poloidal coordinate are obtained for $\tilde{r}=5$ and different $\tilde{R}$. The skyrmion radius is $\tilde{R}_s=1$, it corresponds to $\tilde{\Xi}=-1.15$, see Fig.~\ref{fig:Cn-Xi}(b). }
%     \label{fig:E-vs-X2}
% \end{figure}

Since both $\tilde{\mathcal{F}}^2$ and $\tilde{\kappa}_1$ are independent on $\tilde{X}^2$, the second equation in \eqref{eq:ODE-dimless} is split off and can be formulated in terms of the poloidal angle $\vartheta=X^2/r$:
\begin{equation}\label{eq:pol}
    \partial_{\tilde{t}}\vartheta=\frac{\tilde{r}}{1+\alpha^2\mathcal{C}_0^2}\left\{\tilde{u}^1N_{\text{top}}\left[\mathcal{C}_0(\alpha-\beta)-\frac{\beta\tilde{\mathcal{C}}_1}{\tilde{R}}\frac{\sin\vartheta}{1+\varrho\sin\vartheta}\right]-\frac{\alpha\mathcal{C}_0\tilde{\Xi}}{\tilde{r}\tilde{R}}\frac{\cos\vartheta}{(1+\varrho\sin\vartheta)^2}\right\},
\end{equation}
where $\varrho=r/R$. 

For the special case \underline{$\alpha=\beta$}, Eq.~\eqref{eq:pol} has solution $\vartheta=\vartheta_c=\text{const}$ for any value of the current. Constant $\vartheta_c$ is determined by the equation
\begin{equation}
    \frac{\cot\vartheta_c}{1+\varrho\sin\vartheta_c}=-\tilde{u}^1N_{\mathrm{top}}\frac{\tilde{\mathcal{C}}_1}{\mathcal{C}_0\tilde{\Xi}}\tilde{r}.   
\end{equation}
So, with the increase of current $\tilde{u}^1$, the angle $\vartheta_c$ reaches either $0$ or $\pi$ depending on the direction of $\tilde{u}^1$ and the initial position of the skyrmion. In the limit $\varrho\ll1$, the linear skyrmion velocity can be estimated as
\begin{equation}
    \tilde{V}^1\approx\tilde{u}^1\left[1+\frac{\tilde{\mathcal{C}}_1}{\mathcal{C}_0\tilde{R}}\frac{1}{\sqrt{1+\left(\frac{\tilde{u}^1\tilde{\mathcal{C}}_1\tilde{r}}{\mathcal{C}_0\tilde{\Xi}}\right)^2}}\right].
\end{equation}

In the limit of large currents, one obtains $\tilde{V}^1\approx\tilde{u}^1+\tilde{\Xi}/(\tilde{r}\tilde{R})$.

\begin{figure}
    \centering
    \includegraphics[width=0.8\linewidth]{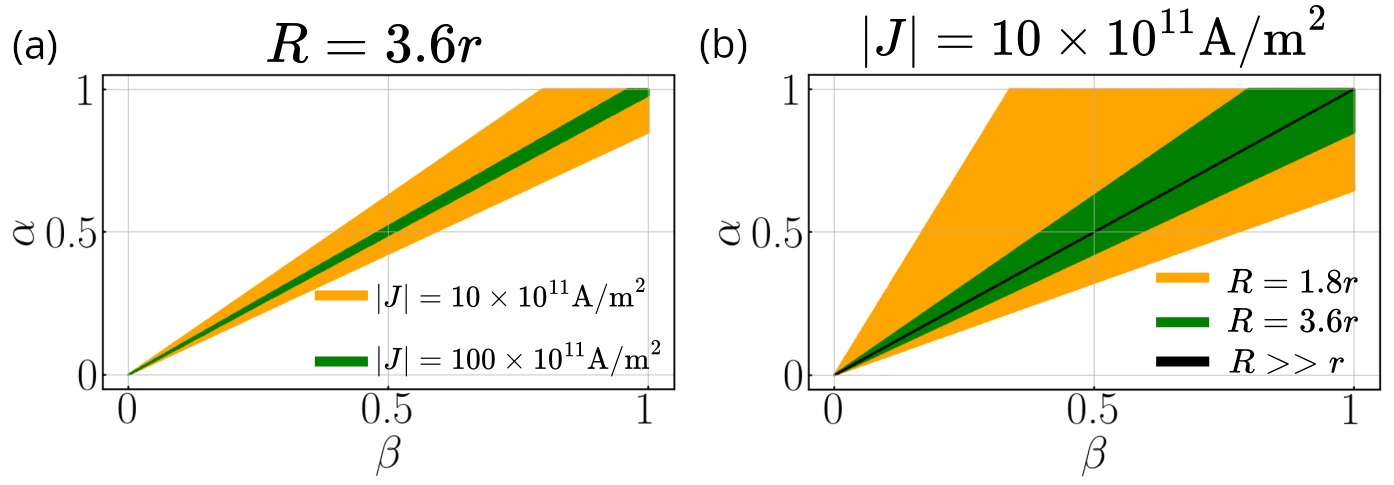}
    \caption{Numerical results of Eq. \eqref{eq:ODE} of the two-dimensional region in the $(\alpha,\beta)$ plane surrounding the line $\alpha=\beta$ for which the translational motion occurs. Panels (a) and (b) depict, respectively, the effects of changing the electric current density and curvature.}
    \label{fig:SM_alphabeta}
\end{figure}

For the general case \underline{$\alpha\ne\beta$}, Eq.~\eqref{eq:pol} can be easily analyzed in the limit $\varrho\ll1$. In this limit, the translational motion $\vartheta=\vartheta_0$ takes place when at least one of the two conditions is satisfied:
\begin{equation}
   |\tilde{u}^1|<\tilde{u}_{\mathrm{w}},\qquad\text{or}\qquad\Upsilon>1. 
\end{equation}
Here
\begin{equation}\label{eq:uc}
    \tilde{u}_{\mathrm{w}}=\frac{\alpha}{|\alpha-\beta|}\frac{|\tilde{\Xi}|}{\tilde{R}\tilde{r}}\frac{1}{\sqrt{1-\Upsilon^2}},\qquad\Upsilon=\frac{\beta}{|\alpha-\beta|}\frac{|\tilde{\mathcal{C}}_1|}{\mathcal{C}_0\tilde{R}}.
\end{equation}
The poloidal position $\vartheta_c$ of the skyrmion is determined by the equation
\begin{equation}
    \cos(\vartheta_c-\varphi)=\frac{\tilde{u}^1}{\sqrt{\tilde{u}_c^2(1-\Upsilon^2)+(\tilde{u}^1\Upsilon)^2}},
\end{equation}
where $\tan\varphi=\tilde{u}^1\Upsilon N_{\text{top}}/(\tilde{u}_c\sqrt{1-\Upsilon^2})$.

In the main text, we analyzed the condition for skyrmion translational motion within the limit $\varrho \ll 1$. This condition defines a two-dimensional region in the $(\alpha,\beta)$ plane surrounding the line $\alpha=\beta$, as shown in Fig.~4(c). To complement this analysis, we now extend the study beyond the $\varrho \ll 1$ regime by numerically solving Eq.~\eqref{eq:ODE} over the full $(\alpha,\beta)$ plane, without approximations. Fig.~\ref{fig:SM_alphabeta} maps the region in the $\alpha$-$\beta$ parameter space where translational motion occurs for $\varphi=2\pi$. Figs.~\ref{fig:SM_alphabeta}(a) and (b) illustrate, respectively, the influence of current density and curvature (varying as a function of $R$) on this region surrounding the line $\alpha=\beta$. These results provide insight into the role of the CCD term in skyrmion dynamics.

\section{Micromagnetic simulations}
\begin{figure}
    \centering
    \includegraphics[width=0.8\linewidth]{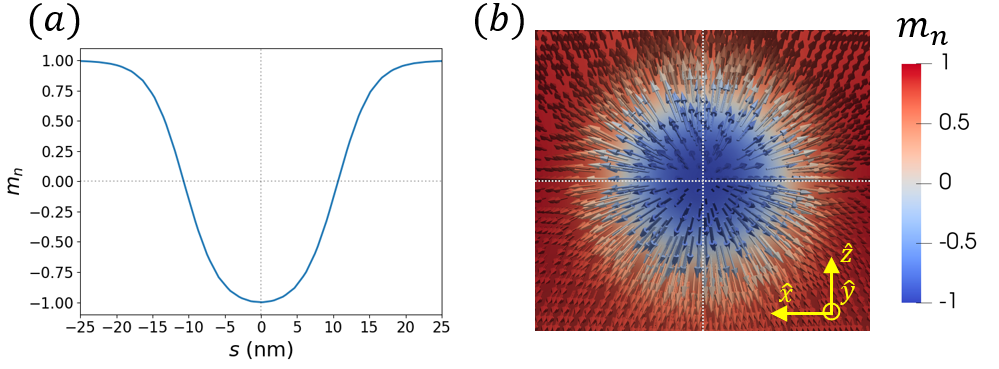}
    \caption{Skyrmion stabilized on a torus with major radius $R=180$ nm, minor radius $r=50$ nm, and thickness  $h = 2$ nm. (a)Azimuthal profile of the normal magnetization component, $m_n$, as a function of the arc-length coordinate $s$, measured from the skyrmion center. (b) Spin texture of the stabilized skyrmion, where the color scale represents the normal magnetization component  $m_n$. }
    \label{fig:skrelax}
\end{figure}

To corroborate the analytical results, micromagnetic simulations were performed using the GPU-accelerated micromagnetic finite-element software Tetmag  \cite{TetMag}, which solves the Landau-Lifshitz-Gilbert equation:

\begin{align}
 \dfrac{d\mathbf{m}}{dt} = - \dfrac{\gamma}{1+\alpha^2}[ \mathbf{m} \times \mathbf{H}_{eff} + \alpha \mathbf{m}\times (\mathbf{m}\times \mathbf{H}_{eff} )]
\end{align}
where $\gamma$ is the effective gyromagnetic ratio, and $\alpha$ is the damping constant. Here $\mathbf{H}_{eff}$ is the effective field that take in account the contribution of the exchange field, anisotropy field, and the interfacial Dzyaloshinskii Moriya field:

\begin{align}
    \mathbf{H}_{eff} = \dfrac{2A_{ex}}{\mu_0 M_s^2} \nabla^2 \mathbf{m} + \dfrac{2 K}{ \mu_0 M_s} (\mathbf{m}\cdot \mathbf{u})\mathbf{u} - \dfrac{2 D}{\mu_0 M_s} [\mathbf{n} (\nabla\cdot \mathbf{m}) - \nabla (\mathbf{m}\cdot \mathbf{n}) ] \, ,
\end{align}
The magnetostatic interaction is incorporated through an effective anisotropy approximation,
$K_{\mathrm{eff}} = K - \mu_0 M_s^2/2$, which is appropriate for thin shells where the thickness is much smaller than the nanotube radius and allows a substantial reduction of computational cost. Since the current version of tetmag does not directly include interfacial DMI, this interaction was implemented by extending the existing bulk-DMI module using the corresponding effective-field formulation.  The implementation was validated through comparison with simulations performed using an open-source micromagnetic  framework \cite{FenicsxMicromagnetics} based on FEniCSx \cite{fenicsxpackage}, showing a very good agreement.\\

The magnetic parameters are taken from the Pt/Co/AlOx structure \cite{Kravchuk-PRL}, i.e, $M_s = 1.09817 \times 10^{6}$ A/m, $A_{ex} = 1.6\times 10^{-11}$ J/m, $K_{eff} = 5.9 \times 10^5$ J/m$^3$, and $D = 2.8$ mJ/m$^2$. The nanotube geometry was meshed using Gmsh \cite{Geuzaine2009} with an average element size of $\sim 1.5$ nm, resulting in approximately $2.2\times10^{6}$ finite elements. The skyrmion configuration was nucleated by initializing the magnetization along the outward surface normal ($\mathbf m=\mathbf n$) and reversing it inside a cylindrical region of radius 15 nm centered at $x=z=0$, in the region $y>0$.  The system was subsequently relaxed for 20 ns with precessional dynamics disabled to obtain a metastable equilibrium configuration. The resulting relaxed magnetization configuration is shown in Figure \ref{fig:skrelax}.

\section{Non-linear regime to the skyrmion motion}

\begin{figure}
    \centering
    \includegraphics[width=0.7\linewidth]{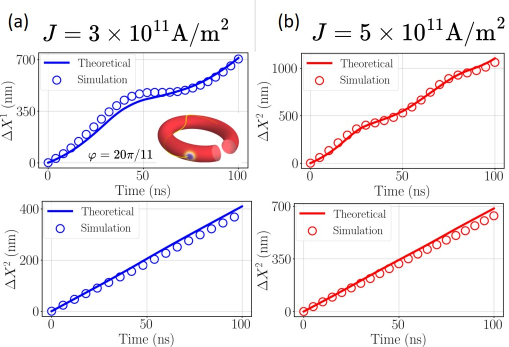}
    \caption{Toroidal (top) and poloidal (bottom) positions of the skyrmion as a function of time for various current density values. In both cases, the skyrmion propagates within the non-linear regime while rotating around the nanotube axis.}
    \label{fig:S10}
\end{figure}

We have analyzed the validity of our model when the skyrmion propagates along the nanotube under the action of electric currents large enough to drive the system into the non-linear regime. In this regime, the skyrmion Hall effect causes the skyrmion to rotate around the nanotube axis while moving forward. This complex dynamics is characterized by time-dependent toroidal (top) and poloidal (bottom) positions, where small upward and backward oscillations in the trajectory arise from energy variations as the skyrmion winds around the tube. The amplitude of this oscillatory motion diminishes as the electric current increases from $3 \times 10^{12}\text{ A/m}^2$ (Figure \ref{fig:S10} a) to $5 \times 10^{12}\text{ A/m}^2$ (Figure \ref{fig:S10} b), while the net toroidal and poloidal velocities increase with the current density.

Throughout both regimes, the analytical model is represented by solid lines, whereas the points denote data obtained via micromagnetic simulations. The excellent quantitative agreement between the analytical curves and the numerical points demonstrates that our model accurately captures the non-linear trajectory and describes the skyrmion motion on curved geometries with high fidelity. This strong match confirms that the rigid skyrmion Ansatz remains robust even when the particle-like texture rotates around the nanotube axis.

%\bibliography{Bibliography}

\end{document}